\documentclass[useAMS,usenatbib,usegraphicx]{mn2e}
\usepackage{amssymb,amsmath}
\title[K+A galaxies ]{The spectral energy distributions of K+A galaxies from the UV
to the mid-IR: stellar populations, star formation and hot dust}

\author[J. Melnick \& R. De Propris]  { J.~Melnick,$^1$ R. De Propris$^2$  \\
   $^1$European Southern Observatory, Santiago, Chile\\
   $^2$Cerro Tololo Inter-American Observatory, La Serena, Chile
   }

\begin{document}

\date{}
\pagerange{\pageref{firstpage}--\pageref{lastpage}} \pubyear{2012}
\maketitle
\label{firstpage}

\begin{abstract}
We present spectrum synthesis fits to 808 K+A galaxies selected from the Sloan Digital 
Sky Survey (SDSS) and population synthesis of their spectral energy distributions, extending
from the far UV (0.15 $\mu$m) to the mid IR (22 $\mu$m), based on the results of 
{\tt STARLIGHT} code fitting to the SDSS spectra. Our modelling shows that K+A galaxies 
have undergone a large starburst, involving a median 50\% of their present stellar masses,
superposed over an older stellar population. The metal abundance of the intermediate-age 
stars shows that star formation did not take place in pristine gas, but was part of a dramatic
increase in the star formation rates for originally gas-rich objects. We find no evidence for 
on-going QSO activity in the UV, which is well modeled by the emission of intermediate-age 
stars. We use K+A galaxies as local counterparts of high redshift objects to test for the 
presence of Thermally Pulsing AGB stars in similarly-aged populations and find no excess 
in the infrared due to emission from such stars, arguing that more distant galaxies
are indeed old and massive at their redshift. All of our galaxies show significant excesses in 
the mid-IR compared to the  light from their stars. We fit this {\it ad hoc} with a 300K
blackbody. Possible sources include TP-AGB stars, obscured young star clusters and 
hidden AGNs, heating a significant dust component.
\end{abstract}


\section{Introduction}

Most galaxies in the Universe fall into two broad classes in the colour-luminosity plane: a `red
sequence' of largely quiescent, typically early-type, old and metal rich systems, or within a 
`blue cloud' of star-forming, later-type, more metal poor and usually less massive objects. 
The bimodality appears to be robust as a function of environment and is observed even at 
higher redshift \citep{Strateva2001,Baldry2006,Brammer2009,Muzzin2012}\footnote{The 
literature on this subject is vast and it is impossible to do justice to the issue within a short
introduction; here we mention a few papers we find especially useful as starting points for 
readers interested in exploring this topic further}. The origin of the bimodal distribution and  
how galaxies migrate between the blue cloud and red sequence (and vice versa) are topics of 
active current research, both observational and theoretical. It is clear that in this context
galaxies in the `green valley', crossing between the two main clumps in colour-magnitude
space, may provide important clues as to the processes leading to the origin of the red
sequence and blue cloud, and how galaxies evolve between them.

In this framework, the class of objects known as K+A galaxies is particularly interesting. 
These appear to be galaxies whose spectrum shows signs of a recent (within 0.2 -- 2 Gyr) 
episode of star formation, producing a strong Balmer series due to A stars, superposed over 
an old stellar population (dominated by K giants) typical of spheroids and early-type spirals.
Nevertheless, there is no strong [OII] or H$\alpha$ emission indicating that star formation is
currently not taking place (e.g., \citealt{Couch1987,Quintero2004}). In many cases, the starburst
appears to have comprised a significant fraction (20--60\%) of the galaxy mass (e.g., \citealt{
Kaviraj2007}; Choi, Goto \& Yoon 2009 and see below). K+A's,  therefore, are observed in a 
short-lived phase as they transit from the blue cloud to the red sequence and thus provide 
the opportunity to obtain unique insights into the processes leading to the morphological 
and spectrophotometric transformation of galaxies.

Major starbursts of the importance observed in these objects are often encountered in 
mergers and interactions (e.g., the Antennae) and there is in fact some morphological 
evidence that a substantial fraction of K+A galaxies may be in interacting systems 
(see e.g. Yamauchi, Yagi \& Goto 2008; \citealt{Yang2008} and references therein). 
While we therefore have some hints as to what triggered the massive starbursts, the real 
mystery is  what made star formation stop so suddenly that today we basically observe no
traces of on-going star formation.  

However, a significant fraction of the current star formation in  distant clusters (where K+A 
galaxies were originally discovered by \citealt{Dressler1983}), is obscured by dust (Duc et 
al. 2002, Saintonge, Tran \& Holden 2008, Dressler et al. 2009) suggesting that  dust 
obscuration may explain the lack of {\em observed} star formation in K+A galaxies.  Shioya, 
Bekki \& Couch (2001) suggest that the progenitors of K+A galaxies may have been dusty 
starbursts of the e(a) type. It is also possible that star-forming is continuing: A-type stars 
could migrate away from their parental  dust clouds during their lifetime, while more massive 
stars would still remain heavily obscured \citep{Poggianti2000}, thus producing an A-rich 
spectrum. \cite{Smail1999} found evidence of on-going star formation in their sample of 5 
K+A's while  \cite{Miller2001}  detected 1.4 Ghz emission in 2 out of 15 K+A galaxies (this 
may be due to an AGN or to synchrotron radiation from supernova remnants in an obscured
star forming region). \cite{Goto2004} set a limit of  $15 M_{\odot}$ yr$^{-1}$ to the current 
star formation in a few K+A galaxies observed with the VLA. At least some K+A galaxies are 
seen to have substantial gas reservoirs comparable to those of spirals of similar luminosities
\citep{Chang2001,Buyle2006}, suggesting that these galaxies may potentially be able to 
reinitiate  star formation in the future. 

In our previous work we have found that a stacked FIRST (Becker, White \& Helfand 1995)
image of 811 K+A galaxies selected from the SDSS Data Release 7 \citep{Abazajian2009,
Goto2007}  shows only weak 1.4Ghz continuum signal, most of which appears to come from 
a subsample of about 5\% of the galaxies that have (5$\sigma$) unambiguous detections in
the FIRST survey (above 0.8 mJy), approximately equally split between AGNs and star-forming
galaxies. The median stack of the remainder of the galaxies is consistent with a purely 
radio-quiet population \citep{Nielsen2012}. This may suggest that most K+A galaxies are
observed in the `off' phase of an activity cycle, and that such galaxies might actually move 
back and forth between the blue cloud and red sequence until their gas reservoirs are finally
exhausted. Such a cycle of activity, modulated by mergers and 'feedback',  has been 
proposed by \cite{Hopkins2008} to account for the formation of a tight  red sequence 
within a hierarchical merger scenario. It is therefore possible that K+A  galaxies may be
`playing possum' rather than having permanently migrated towards the red sequence
following a (merger-induced?)  starburst.

K+A galaxies may also provide clues as to the nature of high redshift objects and their star
formation history. There is a significant population of $1 < z < 2$ galaxies that appear to be 
red, quiescent, and massive. Such systems are difficult to explain within standard $\rm 
\Lambda CDM$ formation scenarios, as they appear to be already well-formed despite the
Universe's relative youth \citep{Yan2004, Cirasuolo2008}. \cite{Maraston2005} proposed 
that these objects 'fake' their old ages because they are dominated by $\sim 1$ Gyr old stars, 
which produce Thermally Pulsing AGB stars (TP-AGB), whose high luminosities and red colours 
mask the underlying blue light and therefore yield apparently larger stellar masses and older 
ages \citep{Tonini2009, Tonini2010}, even for young and relatively low mass objects.  
Present-day K+A galaxies of course also host older stellar populations, but they have  
experienced a major  star-formation episode, involving a large fraction of their stellar mass, 
and may therefore be  adequate local counterparts to test this hypothesis \citep{Kriek2010,
Zibetti2012}, whereas detection of the spectral signatures  of TP-AGB stars would be very 
difficult at high redshifts. 

In this paper we present a statistical study of the properties of the 811 K+A galaxies studied
by \cite{ Nielsen2012}, for which we have combined SEDs obtained using archival data from
GALEX, SDSS, 2MASS and WISE, with the population synthesis study of the SDSS spectra by
\cite{Cid2005}. We use the combined knowledge of stellar populations, models, and spectral
energy distributions covering the entire thermal spectrum over 2 decades in wavelength (from
0.15 to 22 $\mu$m) to discuss the internal constituents of K+A galaxies, test current star
formation and the nature of the starburst, and compare with high redshift galaxies. In the 
next section we describe the dataset, photometry and spectroscopy, and population 
synthesis models. We then discuss and interpret the results and present our conclusions 
and suggestions for future work. Unless otherwise stated and where applicable, we use the
'concordance' cosmological parameters derived from the latest analysis of WMAP data
in \cite{Jarosik2011}. 

\section{Description of the Dataset and Broad Properties}
\label{section2}

We used a sample of 811 K+A galaxies selected from the SDSS as described by 
\citep{Nielsen2012}. This is an expansion of the K+A sample used by \cite{Goto2007}
and drawn from the 7$^{th}$ data release of the SDSS \citep{Abazajian2009}. Briefly, only 
objects classified as galaxies with a spectroscopic $S/N >10$ per pixel are considered. 
The selection criteria of K+A galaxies are equivalent widths of H$_{\alpha}>-3.0$\AA; 
H$_{\delta}>5.0$\AA;  and [OII]$>-2.5$\AA\  where emission lines are negative. Galaxies 
with redshift $0.35 < z < 0.37$ have been excluded due to contamination by the 5577\AA\
 atmospheric feature. 

\begin{figure}
\includegraphics[width=0.5\textwidth, height=2.9in]{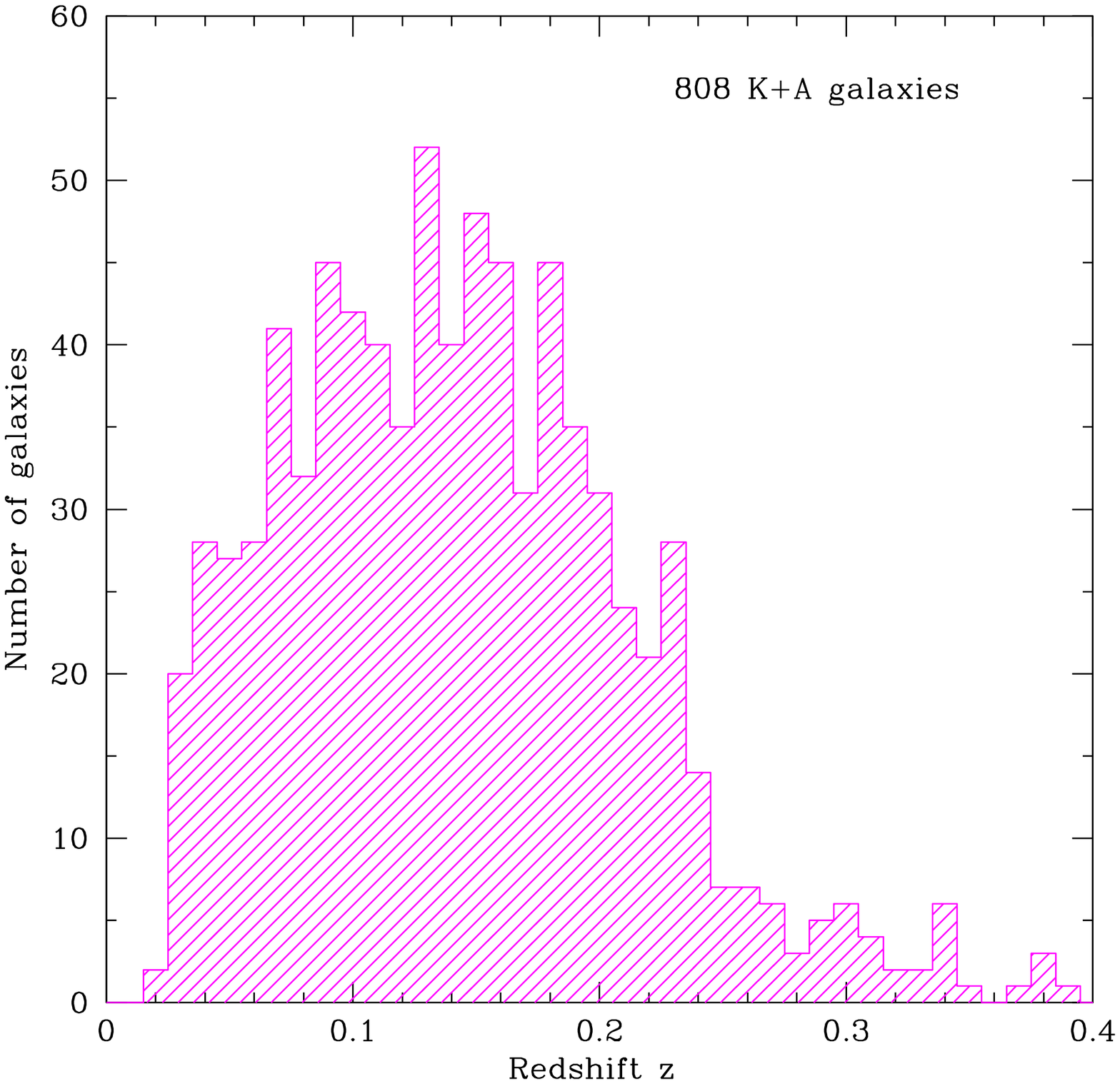}\\
\includegraphics[width=0.5\textwidth, height=2.9in]{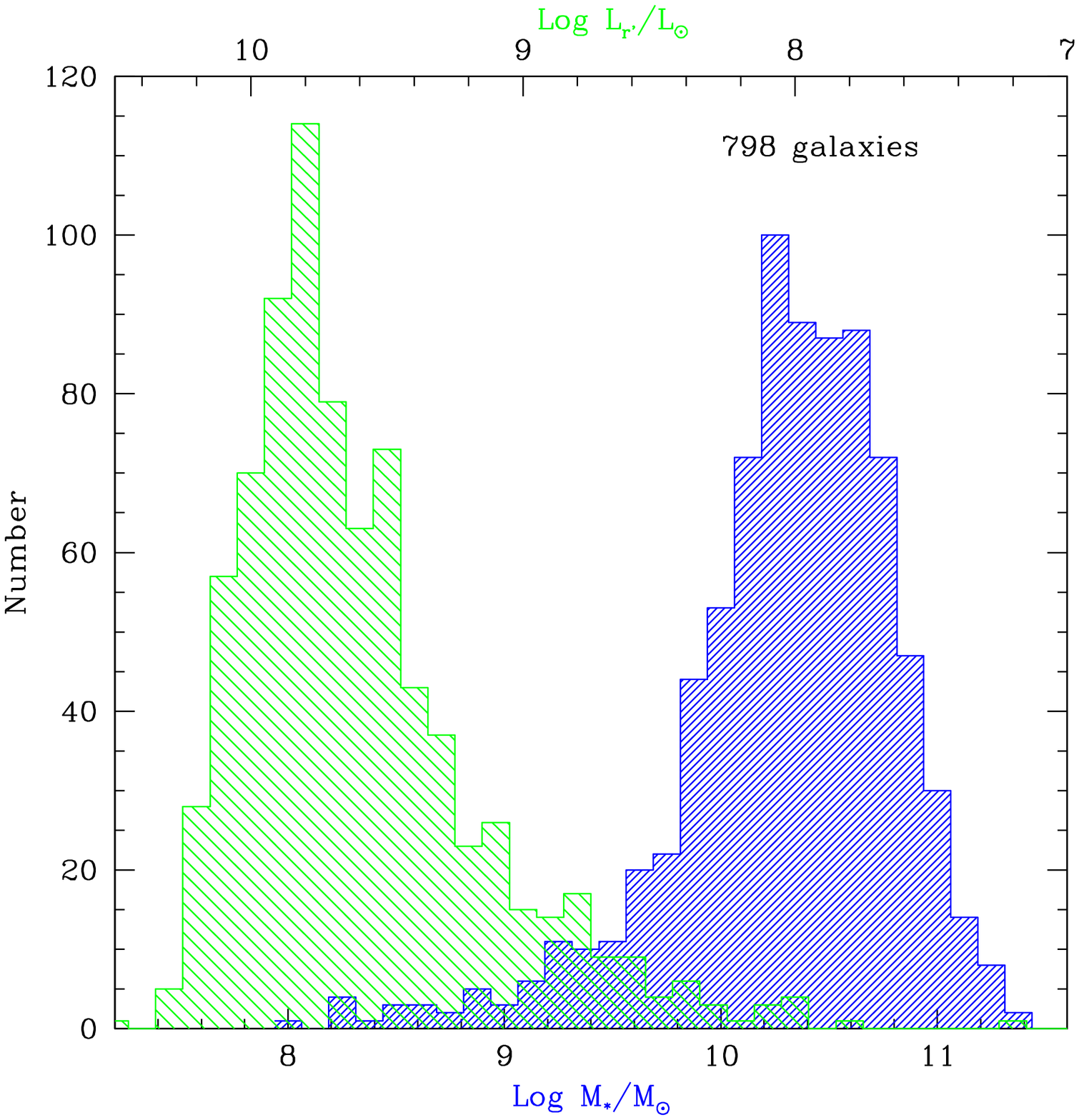}\\
\includegraphics[width=0.5\textwidth, height=2.9in]{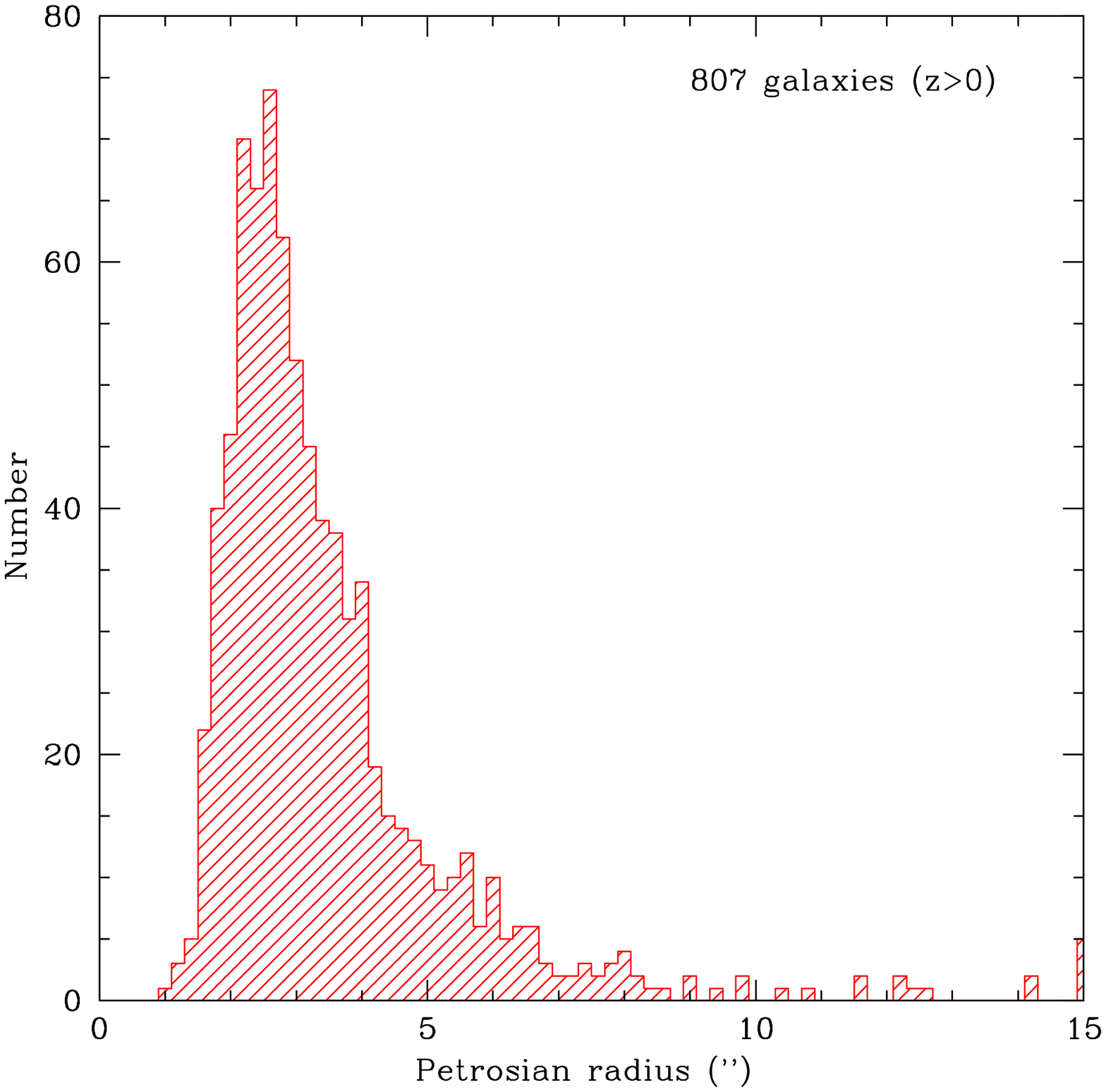}
\caption{Properties of the K+A sample used in this paper. From top to bottom we show the
redshift distribution, distribution of stellar masses from {\tt STARLIGHT} (see below) in solar units, the distribution of $r$ band luminosities in solar units,
and the angular size distribution in seconds of arc as characterized by the Petrosian radii returned by the SDSS pipeline.}
\label{sample}
\end{figure}

Figure~\ref{sample} summarizes the principal characteristics of the sample. Galaxies span
the redshifts $0 < z < 0.4$ with a median of $z=0.138$ and have typical luminosities of 
$\sim10^{10} L_{\odot}$ and total stellar masses (as given by {\tt STARLIGHT}; \rm see below) of
$\sim3\times10^{10}M_{\odot}$\ comparable to approximately $L^*$ galaxies, with a longer tail towards
lower mass objects. Most objects are resolved and have apparent radii of a few arcsec, 
typical of galaxies of their luminosity at the appropriate redshift. 

Of the 811 galaxies in the \cite{Nielsen2012} sample, we have retrieved spectra from the SDSS
database \citep{Abazajian2009}, as well as the associated SDSS photometry in $ugriz$. 
For these galaxies we then proceeded to obtain multi-wavelength photometry in the far
ultraviolet from the GALEX archive \citep{Morrissey2007}, infrared ($JHK$) data from the
2MASS \citep{Skrutskie2006} and the latest release of the UKIDSS survey\footnote{The 
UKIDSS project is defined in Lawrence et al. (2007). UKIDSS uses the UKIRT Wide 
 Field Camera WFCAM (Casali et al. 2007).  The photometric system is described in 
 Hewett et al. (2006), and the calibration is described in Hodgkin et al. (2008). The 
 pipeline processing and science archive are described in and Hambly et al (2008)}, and finally fluxes 
at 3.4, 4.6, 12 and 22 $\mu$m from the WISE dataset \citep{Wright2010}. We also 
used archival data from the Spitzer Space Telescope for a few galaxies observed as 
part of other programs (e.g., Lockman Hole, ELAIS).

Given the fact that this study is entirely based on existing archival data, from the outset 
we were concerned about possible systematic effects due to aperture matching in the
photometry. Thus, we carefully inspected the relevant data bases for the measurements
that would minimize any systematic difference for the different catalogues. 

Our base photometry is the optical $u,g,r,i,z$ from the SDSS, which is measured in Petrosian
apertures that are expected to approximate the total flux of galaxies. GALEX photometry is
calculated in similar Kron-style apertures, but the resolution ($1.5''$) and point spread
function ($4.5''$ in the near UV and $6.0''$ in the far UV) are such that, given the sizes of
most our objects (Figure~\ref{sample}), they are basically star-like as seen by this telescope.

The 2MASS survey provides isophotal and extrapolated magnitudes for objects in the Extended
Source Catalog (XSC), comprising 291 of our 811 K+A galaxies. Comparison with UKIDSS 
Petrosian magnitudes (for 71 of these objects which are in common with the XSC) showed that
the isophotal estimates are more stable and closer to a `SDSS-style' total magnitude, and
these were adopted. Almost all of our galaxies are also in the 2MASS Point Source Catalog
(PSC). For the remaining 620 objects (in the PSC) we also carried out a comparison with
the UKIDSS photometry in the appropriate bands and eventually concluded that the 2MASS
$4''$ aperture magnitude yields a closer estimate to the total magnitude and this was 
therefore employed in our analysis (note that this is comfortably larger than the Petrosian 
sizes of most of our sample as shown in Figure~\ref{sample} above).

WISE gives magnitudes within the XSC isophotal apertures if objects are positionally matched
to 2MASS and `adaptive' apertures for all other galaxies. We accepted the WISE estimate of the
best magnitudes for our galaxies, as provided by their database. As it turned out, 12 of our 
objects have been observed by IRAC and MIPS24 with SPITZER. The corresponding SPITZER 
Kron magnitudes compare very well with the WISE values in the $3.4-3.6\mu m$ and
$4.5-4.6\mu m$ bands, which gives us confidence that the WISE magnitudes are measured
through apertures consistent with the rest of the bands.

Table~\ref{tab1} shows the first 10 objects and their photometry; the full table is available 
on the MNRAS website in machine-readable format. Only for 98 galaxies in our
sample we have full FUV(0.15$\mu m$) to MIR (22$\mu m$) SEDs; while for essentially 
the full sample we have $u'$ to $12\mu$ data. 

\begin{table*}
\tabcolsep 2.0mm
\tiny
\centering 
   \caption{\bf Spectral Energy Distributions ($\times 10^{-17} erg s^{-1} cm^{-2} \rm \AA^{-1}$)}                                                   
   \begin{tabular}{@{}  l  c  c  c  c  c  c  c  c  c  c  c  c  c  c  c c c@{}}           
   \hline\hline                   
&                  & \multicolumn{2}{c}{\bf GALEX}  &  \multicolumn{5}{c}{\bf SDSS} & \multicolumn{4}{c}{\bf 2MASS or UKIDSS} &  \multicolumn{4}{c}{\bf WISE} \\
&   Galaxy  &               FUV & NUV               &            u' & g' & r' & i' & z        &        Y &   J & H & Ks                     &  $3.4\mu$ & $4.6\mu$ & $12\mu$ & $22\mu$ \\ 
\hline
      1& 001027.392-104341.71&   0.00&  14.57&  21.91&  46.73&  49.51&  41.73&  35.40&   0.00&  21.21&  10.78&   7.50&   1.83&   0.93&   0.55&   0.39 \\ 
   2& 001551.253-102317.92&  17.01&  11.74&  13.21&  35.22&  38.24&  31.69&  26.42&   0.00&  16.41&  10.58&   5.92&   1.14&   0.49&   0.07&   0.00 \\ 
   3& 010823.370-094356.22&   0.00&   0.00&   8.33&  17.70&  24.44&  22.44&  19.46&   0.00&  12.44&  11.31&   5.15&   1.13&   0.44&   0.00&   0.19 \\ 
   4& 011942.227+010751.63&  16.51&  24.61&  25.91&  56.62&  57.04&  48.34&  42.05&   0.00&  32.57&  24.67&  11.38&   2.39&   0.81&   0.08&   0.00 \\ 
   5& 012015.683-095920.12&   0.00&  10.17&  12.92&  34.19&  37.75&  32.13&  27.35&   0.00&  22.10&  12.20&   3.63&   1.62&   0.60&   0.07&   0.24 \\ 
   6& 014327.340-011342.55&   3.69&   6.66&   7.75&  20.41&  28.80&  26.64&  22.75&  20.55&  16.15&  11.84&   7.26&   1.47&   0.64&   0.35&   0.37 \\ 
   7& 014838.633-093259.77&   3.90&   0.00&   9.12&  24.25&  32.38&  28.64&  24.63&   0.00&  16.19&  10.64&   6.16&   1.29&   0.49&   0.07&   0.00 \\ 
   8& 015107.018-005636.72&   1.49&   4.93&   9.69&  24.30&  31.78&  28.20&  24.74&  21.92&  18.45&  12.27&   6.95&   1.59&   0.65&   0.11&   0.00 \\ 
   9& 020505.984-004345.25&  23.26&  29.85&  27.71&  47.72&  46.78&  39.83&  31.64&  26.33&  19.83&   9.36&   5.98&   1.18&   0.44&   0.15&   0.00 \\ 
  10& 021007.624-095431.19& 227.70& 181.10&  79.85& 126.30& 105.20&  85.36&  62.76&   0.00&   0.00&   0.00&   0.00&   1.19&   0.39&   0.17&   0.11 \\ 
 \hline
   \end{tabular}
   \label{tab1}
\end{table*} 

The table shows the SDSS name (in the form hhmmss.sss and ddmmss.ss) and the fluxes in 
GALEX (FUV and NUV), SDSS ($ugriz$), 2MASS ($JHK$) or UKIDSS ($YJHK$), and the four WISE bands (in this order) 
in units of $10^{-17}$ ergs cm$^{-2}$ s$^{-1}$ \AA$^{-1}$.  Upper limits are shown as
zeros in the table. For reasons of clarity and economy of space, Table~\ref{tab1} does not
give the measurement errors, which are publicly available from the source web-sites. It should
be reiterated here that the magnitudes come from a variety of methods on data and
instruments of varying resolution and therefore we assume that these individual
measurements adequately approximate the total magnitude of galaxies without distorting 
their spectral energy distributions (this amounts to saying that the apertures used are
reasonably similar and colour gradients weak). The tabulated fluxes have been corrected 
for extinction as described below.

We will first analyze the stellar populations of these galaxies based on modelling the
SDSS spectra and line ratios. This information will then be used to inform our understanding 
of their spectral energy distributions and the nature of the astrophysical sources contributing
to the light in different wavelength regimes.

\subsection{Stellar Populations from Spectrum Synthesis Models}

\cite{Stasinska2008} have used {\tt STARLIGHT} to fit simple stellar population (SSP) models to 
hundreds of thousands of galaxy spectra from the SDSS/DR7 data release.  Full details about 
the modelling process are given by \cite{Cid2005}. Briefly, the spectral fits are based on the 
Bruzual \& Charlot (2003; BC03) models with the STELIB library \citep{LeBorgne2003}, and the 
Padova 1994 evolutionary tracks \citep{Bertelli1994} with a \cite{Chabrier2003} IMF.  The data, 
the model fits for the galaxies in our sample, and the code itself are publicly available from the {\tt STARLIGHT} 
website,\footnote{http://www.starlight.ufsc.br} and many details about the use of these models 
for studies of galaxies with very weak emission lines are given in the series of papers by the 
SEAGal collaboration cited by \cite{Cid2011}.

\begin{figure*}
\hspace*{-0.1cm} \includegraphics[height=9cm,width=9cm]{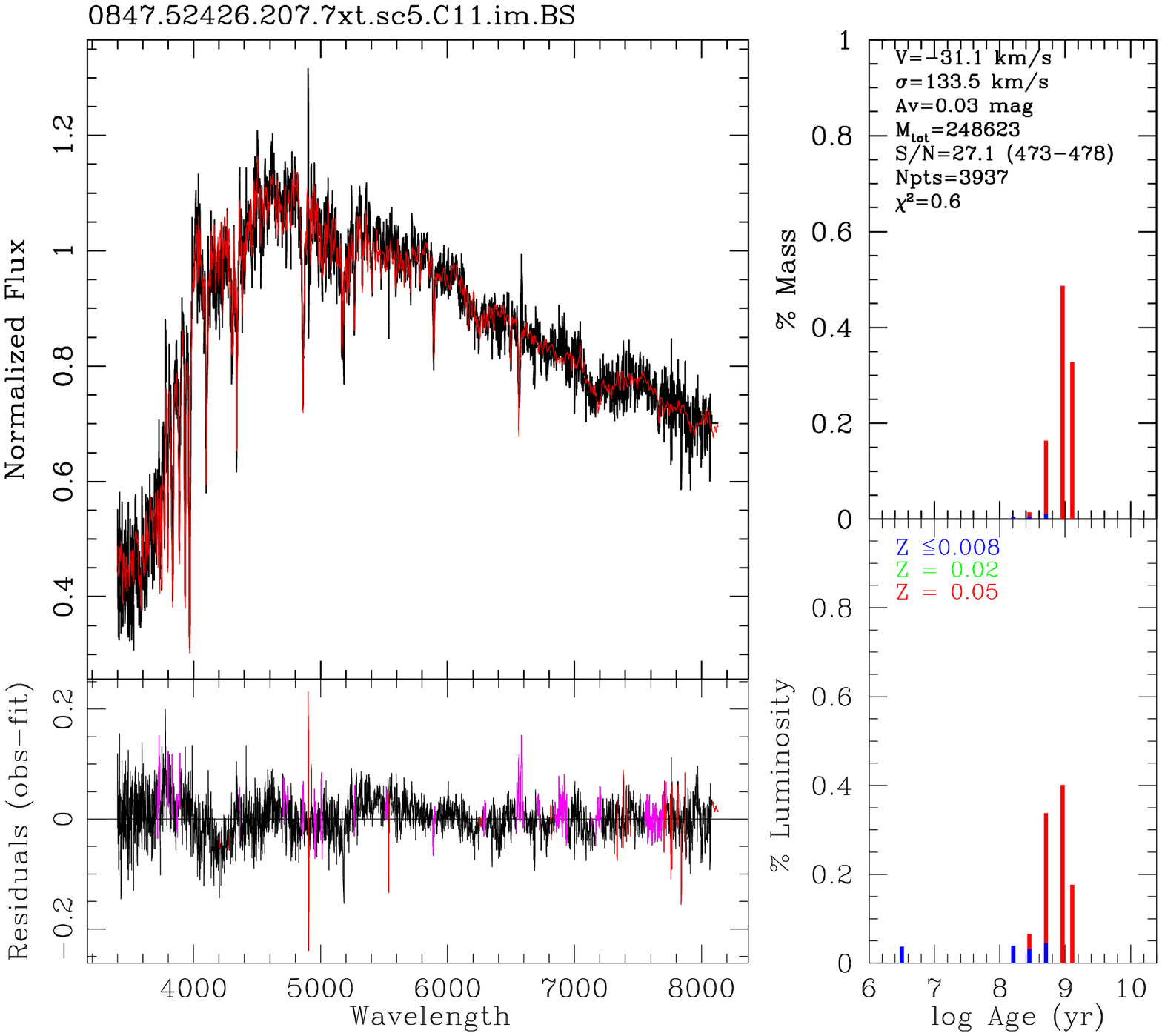}\hspace*{0.0cm}\includegraphics[height=9cm, width=9cm]{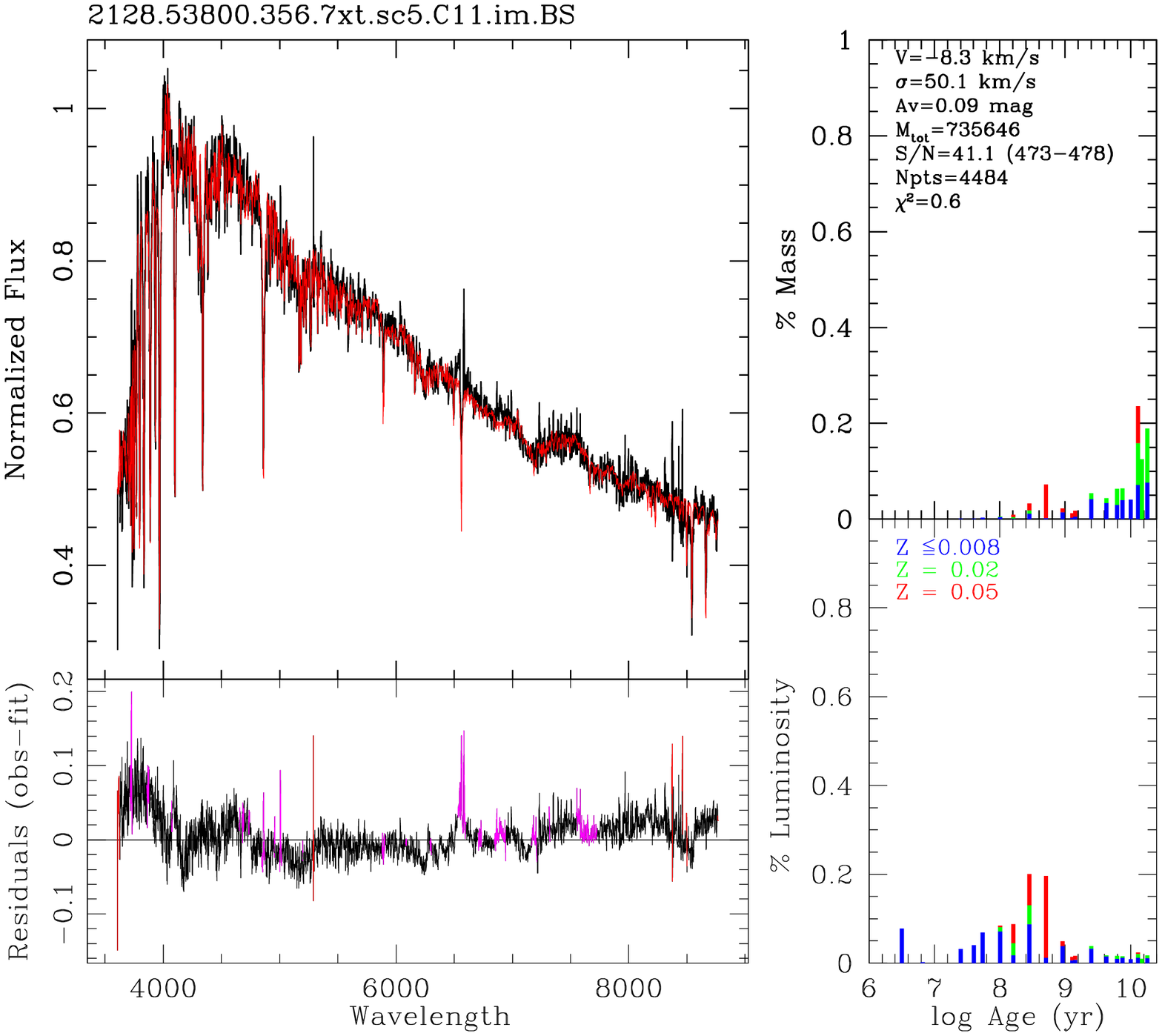}
\caption{{\bf Left.} Example of {\tt STARLIGHT} modelling for a galaxy where both the luminosity
and the stellar mass are dominated by the A-component.  The larger panels show the 
observed (black) and fitted (red) spectra and the residuals, which clearly show the emission lines (that 
have been masked from the population fit so they appear in purple). The smaller panels 
show the fraction of the luminosity as a function of age for 3 different metallicities $Z$ (bottom), 
and the fraction of the stellar mass  plotted in the same way on the top panel. The colour-coding of 
metallicities (blue: metal poor -- $Z\leq0.5Z_{\odot}$; green: solar; red: metal rich --  $Z=2.5Z_{\odot}$) is indicated in the lower panel, while a number of parameters relevant for the 
fits are displayed in the upper panel. Notice that the EW(H$\alpha$) in absorption is larger 
than that in emission, illustrating the importance of subtracting the models in order to derive
meaningful emission-line fluxes. {\bf Right}. Similar to the left panel, but for a galaxy 
with a predominant old population by mass. In this example the emission 
lines are also heavily contaminated by underlying absorption. }
\label{ssp}
\end{figure*}

The SDSS {\tt STARLIGHT} database includes 808 of our 811 galaxies. Therefore, the 
three missing objects (J103833.180+034930.29; J143852.706+613020.53; and
 J144332.436+304807.40),  for which we do have the SEDs and spectra, will not be 
included in our study. The SSP models fit the stellar populations using 25 different ages 
and 6 metallicities.  Examples of {\tt STARLIGHT}  fits for two `typical' K+A galaxies are 
shown in Figure~\ref{ssp} that presents the observed and fitted spectra, the residuals, 
and the star formation histories resulting from the fit (ages, metallicities).

\begin{table}
\tabcolsep 2.0mm
\begin{centering}
\caption{Age and Metallicity bins used to display the results of the SSP model fits. }
\begin{tabular}{| l l l |}
 \hline\hline
 Bins			& {\tt STARLIGHT} 			& {\tt MARASTON-mix} 		\\ 
 			& BC03+Chabrier 			& rhb+Kroupa 			  	\\ \hline 
			&\multicolumn{2}{c}{\hspace*{-0.9cm} Age $t$ [Gyr]}						\\ 
   {\bf young} 		& $t<0.1$					& $t<0.1$				\\
   {\bf intermediate}    & $0.25\leq t \leq 4.25$		& $0.3\leq t \leq 4.25$	\\
   {\bf old}			& $ t>4.25$				&  $t>4.25$			\\ \hline
           		& \multicolumn{2}{c}{\hspace*{-0.9cm}Metallicity $Z$}	 	\\ 
   {\bf metal poor}	& $ Z\leq0.008$			&$Z\leq0.01$ 			\\
   {\bf solar}		& $Z=0.02$ 				&$Z=0.02$ 			\\
   {\bf metal rich}	& $Z=0.05$				&$Z=$(0.04+0.07)/2  	\\ \hline  
   \end{tabular}
 \label{mix}
\end{centering}
\end{table}

\begin{table*}
\tabcolsep 2.0mm
\tiny
\centering 
   \caption{\bf {\tt STARLIGHT} Populations for K+A galaxies}                                                   
   \begin{tabular}{@{}  l c c c c c c c c c c c c c c c c c  @{}}   
   \hline\hline                   
&  SDSS  & $z$ & Mass                     & \multicolumn{2}{c}{Av [mag]}  &    &	\multicolumn{3}{c}{Equivalent widths [\AA]}   &  Young& \multicolumn{3}{c}{Intermediate age  [\%]} &  \multicolumn{3}{c}{Old [ \%] }\\
&               &    &  $10^9$M$_\odot$  & Int. & MW   &  $\rm \frac{[NII]}{H\alpha}$           &               H$\alpha$ & [OIII]  & [OII]                  &[\%]  & poor & solar & rich  & poor  & solar & rich \\  
 
   \hline\small
      1& 001027.392-104341.71&  0.1329&  20.9&  0.32&  0.08&   1.2&   4.9&   7.6&   2.3&   0.0&  23.4&   0.5&  37.5&   8.4&   0.0&  30.2 \\ 
   2& 001551.253-102317.92&  0.1982&  26.9&  0.38&  0.10&   1.0&   1.1&  -0.2&   1.4&   0.2&   3.7&   0.0&  85.4&   0.1&  10.2&   0.0 \\ 
   3& 010823.370-094356.22&  0.1834&  40.0&  0.07&  0.17&   1.4&   0.7&   0.4&   0.4&   0.1&   5.0&   0.3&  30.2&  10.1&  46.5&   7.7 \\ 
   4& 011942.227+010751.63&  0.0900&  24.9&  0.05&  0.05&   0.3&   1.8&  -0.2&   1.9&   0.2&   7.7&   0.9&   6.1&   3.7&  28.8&  52.0 \\ 
   5& 012015.683-095920.12&  0.1369&  25.5&  0.03&  0.14&   2.5&   0.6&   0.2&   1.5&   0.4&  12.9&   0.9&  37.8&  11.6&  35.5&   0.0 \\ 
   6& 014327.340-011342.55&  0.1963&  59.7&  0.20&  0.04&   1.5&   4.2&   5.2&   2.1&   0.1&   5.5&   0.4&   9.7&  26.9&  28.6&  27.8 \\ 
   7& 014838.633-093259.77&  0.2170&  29.1&  0.16&  0.05&   1.8&   0.8&   0.3&   3.1&   0.4&  21.0&   0.0&  73.1&   0.0&   2.6&   0.0 \\ 
   8& 015107.018-005636.72&  0.1981&  29.7&  0.10&  0.04&   1.3&   0.9&  -0.2&   0.5&   0.0&  18.3&   1.2&  68.9&  11.6&   0.0&   0.0 \\ 
   9& 020505.984-004345.25&  0.1134&  16.8&  0.56&  0.01&   0.6&   4.5&   0.9&   3.5&   0.2&   8.4&   0.3&  25.6&  34.6&  20.8&   9.4 \\ 
  10& 021007.624-095431.19&  0.0412&   1.9&  0.34&  0.01&   0.4&   4.2&   0.5&   4.6&   0.1&  12.7&   0.0&   3.6&  81.6&   2.1&   0.0 \\ 
 \hline
   \end{tabular}
   \label{tab2}
\end{table*}
For a variety of reasons, obtaining reliable estimates of the statistical uncertainties in the 
stellar populations derived from ssp model fits is a notoriously difficult task 
(see e.g. \cite{GonzaCid2010} for a recent discussion) and in particular {\tt STARLIGHT} 
does not provide statistical errors.  In a previous investigation \citep{Melnick2012} we
found that using wider age and metallicity bins increases the robustness of the models 
against statistical and systematic (i.e. variations in the input parameters) fluctuations.
Therefore, we  binned the populations into 3 broad metallicity ($Z$) bins and 3 age groups  ($t$) 
as indicated in Table~\ref{mix}. The binned populations  of the first 10 galaxies are listed
 in Table~\ref{tab2}  (the full table is available on the MNRAS website in machine-readable
 format). 

Figure~\ref{popi} summarizes the star formation histories of K+A galaxies as derived from
the {\tt STARLIGHT} fits:

\begin{figure}
\includegraphics[width=0.5\textwidth]{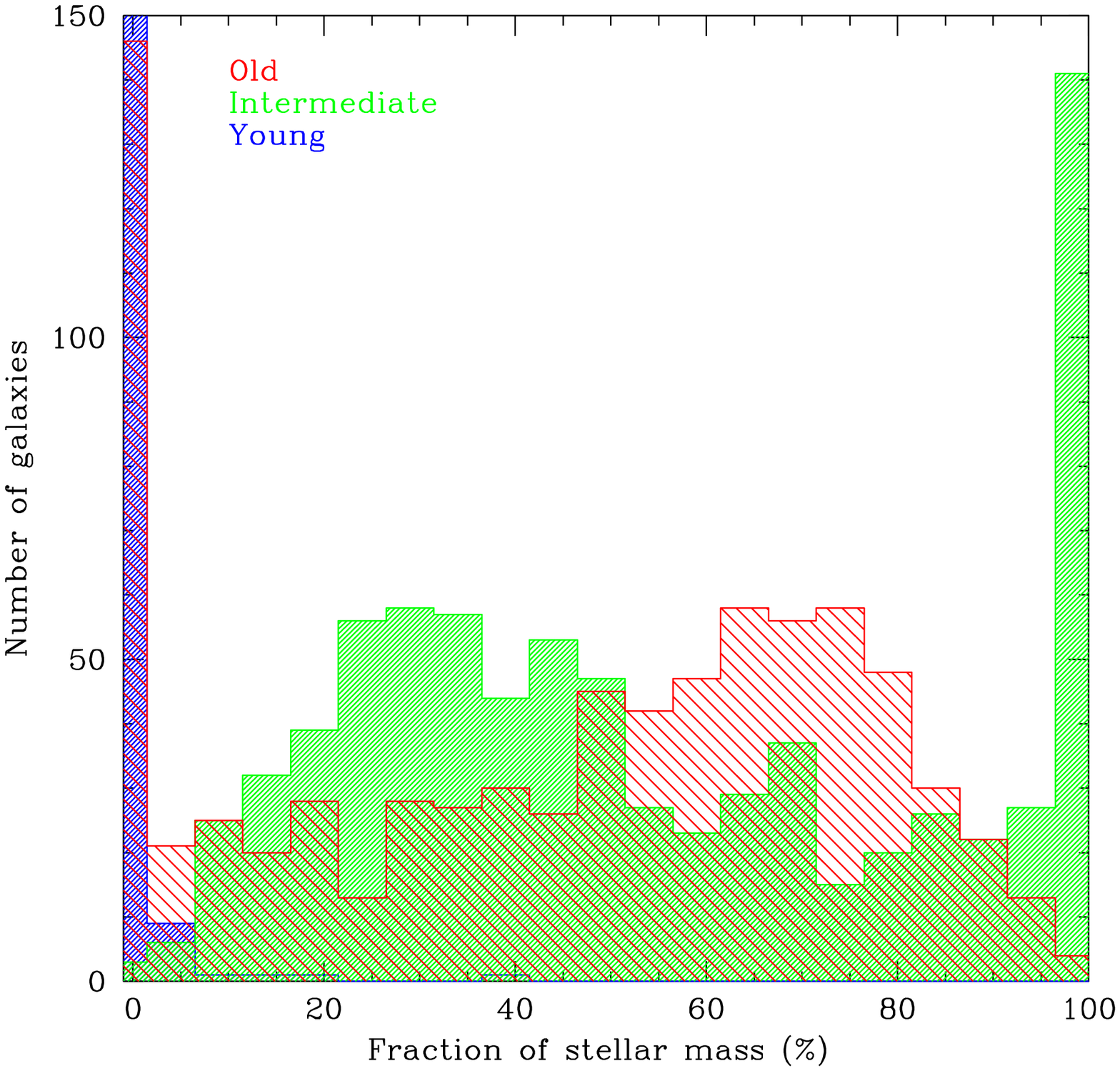}\\
\includegraphics[width=0.5\textwidth]{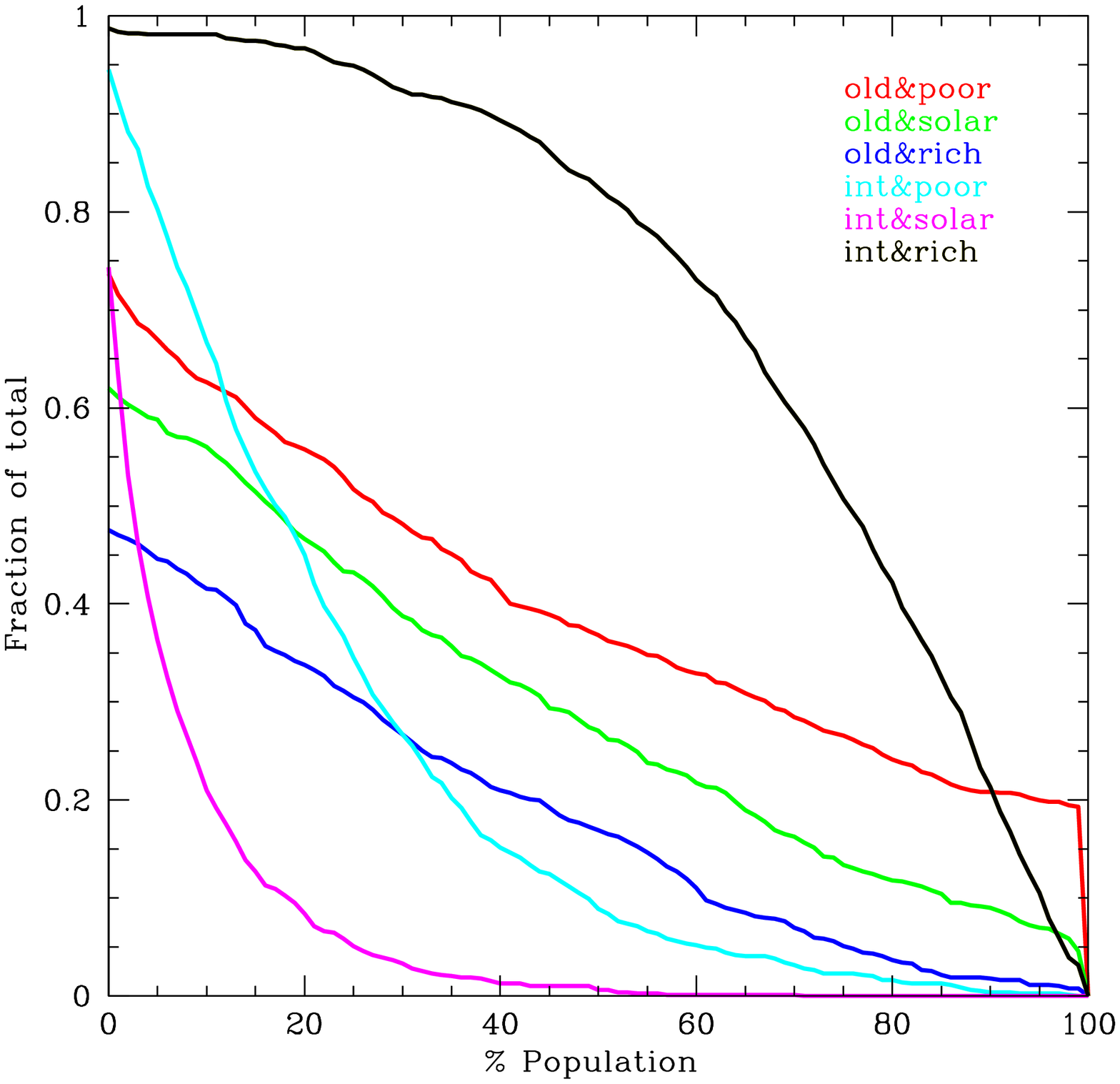}
\caption{Stellar populations by mass in K+A galaxies.  {\bf Top.} The histograms show the distribution of young, intermediate age, and old stars in 
K+A galaxies. The sample contains only one galaxy with a significant young stellar population; intermediate-age and old stars are the dominant populations
{\bf Bottom.} Cumulative fractions of old and intermediate-age stars in the three metallicity bins discussed in 
the text. Metal rich stars dominate the intermediate-age populations, while metal poor stars are the most common among the old stars. The metallicity distribution of
the intermediate-age population is heavily skewed towards metal-rich and metal-poor stars, while solar abundance stars are relatively rare.  }
\label{popi}
\end{figure}

\begin{itemize}

\item On average, the intermediate age stars (formed in the recent star formation episode 
and contributing to the A spectrum) account for close to 50\% of the total stellar mass 
in K+A galaxies. This strongly suggests that the starburst has involved a very significant
portion of the initial mass of the galaxy  \citep{Kaviraj2007,Choi2009}. Analysis of colour gradients
for K+A galaxies by \cite{Yang2008} and IFU spectroscopy by \cite{Pracy2012} argues
that the starburst tended to take place preferentially in the central regions of galaxies.

\item In agreement with the lack of strong [OIII] and H$\alpha$ emission, the population 
of young ($ < 0.1$ Gyr) stars is negligible; the mean of the sample is  $\sim$0.3\% and 
only three galaxies have more than 10\% young stars.

\item The intermediate-age (IA) populations  tend to be predominantly metal poor or 
metal-rich; the median galaxy has close to 80\% metal-rich IA stars, while less than 
10\% of the galaxies have more than 20\% IA stars of solar metallicity. This supports
our previous point that a rapid starburst took place, involving a large fraction of the 
pre-existing gas. Minor mergers with gas-rich subsystems are unlikely to provide
sufficient material to account for the large fraction of younger stars, and would presumably
dilute the existing material with more pristine gas.

\item We suggest that K+A galaxies must have therefore experienced a dramatic increase
in their star formation rates followed by rapid quenching \citep{Brown2009}, given the lack
of current star formation. The abundance pattern observed supports this hypothesis, as we
would expect a gentler and more continuous distribution if star formation had been slowly
throttled by mechanisms such as 'strangulation' or `harassment'.

\item About 15\% of the galaxies do not contain an old stellar population. Otherwise, 
the old stellar population is predominantly metal poor. Close to 50\% of the galaxies 
do not contain old, metal rich stars, while the median fraction of solar abundance stars is
20\%.  The original galaxies must therefore have appeared as late(r)-type spirals of
apparently lower stellar mass. \cite{Mendel2012} suggest that the K+A phenomenon
is related to the growth of the galactic bulge, and therefore morphological transformation.

\end{itemize}

\subsection{Emission-line Diagnostic Diagrams}

Almost all the K+A galaxies in our sample show weak but clearly visible emission-lines of
[NII]6548,6584 and H$\alpha$, and many galaxies also show [OII]3727,3729 and [OIII]5007.  
It is possible, therefore, to use emission-line diagnostic diagrams to gain further insight 
into the stellar populations by exploring the photoionization properties of the sample. 
The SEAGal collaboration (\citealt{Cid2011}; henceforth CFS11 and references therein) have 
extensively explored this issue on a sample of 700 thousand galaxies from the SDSS with 
weak emission lines. In particular, they noted that the traditional  `BPT'  diagnostic diagrams 
that make use of the [OII], [OIII], and H$\beta$ lines (among others) are not suitable for
galaxies with weak emission lines because these lines are usually very weak and, in the 
case of H$\beta$, are heavily contaminated by underlying absorption. This is particularly
important for K+A galaxies where the emission lines are, by definition, weak while Balmer
absorption lines are strong.
 
\begin{figure}
\hspace*{-0.0cm} \includegraphics[width=0.5\textwidth]{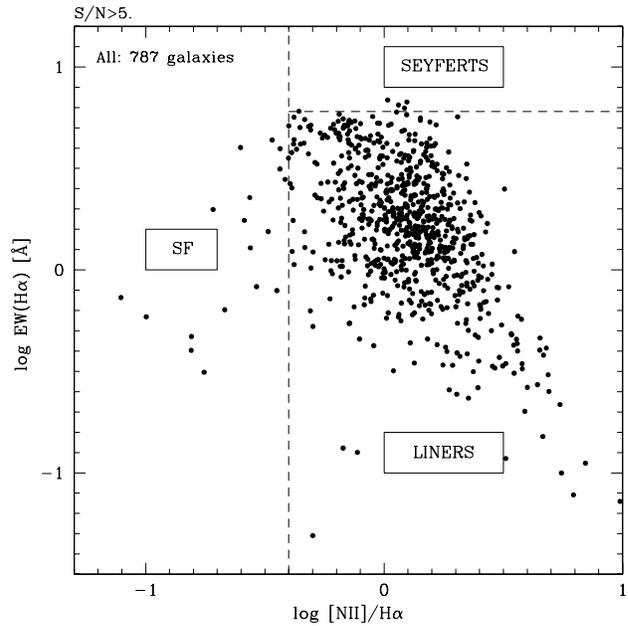} 
\caption{WHAN diagram for 787 K+A galaxies with $\rm S/N>5$ around 660nm.
The traditional boundaries separating LINERs, SEYFERTs, and Star-Forming (SF) galaxies are
shown by the dotted lines (see text for more details). }
\label{EWHa}
\end{figure}

They therefore introduced a new diagnostic diagram, which they call `WHAN', that only 
uses [NII]6584 and H$\alpha$, which are usually the strongest lines in these types of 
galaxies. \cite{Cid2010} and CFS11 have in fact shown  that their EW(H$\alpha$) vs.
 [NII]/H$\alpha$ `WHAN'  diagram retains the diagnostic power of the more traditional 
BPT plots, but also allow to separate `true' from `fake' AGNs as we shall see below,
while being less sensitive to reddening. 

Figure~\ref{EWHa} shows  the WHAN diagram for the 787 galaxies in our sample that 
have $S/N>5$  in the continuum around [NII]  \& H$\alpha$ (as indicated in the top-label 
of the figure).  The dashed lines separate the regions populated by LINERS, SEYFERTS, and 
star-forming galaxies (SF) as defined by \cite{Kewley2006} transported to the WHAN plane 
by \cite{Cid2010}.  The plot shows that, as expected, the overwhelming majority of our 
K+A galaxies fall in the LINER domain. 
   
\subsection{Spectral Energy Distributions (SED) from 0.15 to 22 $\mu$m}
\label{ZEDS}
 
\begin{figure*}
\hspace*{-0.2cm}\includegraphics[height=9cm,width=9cm]{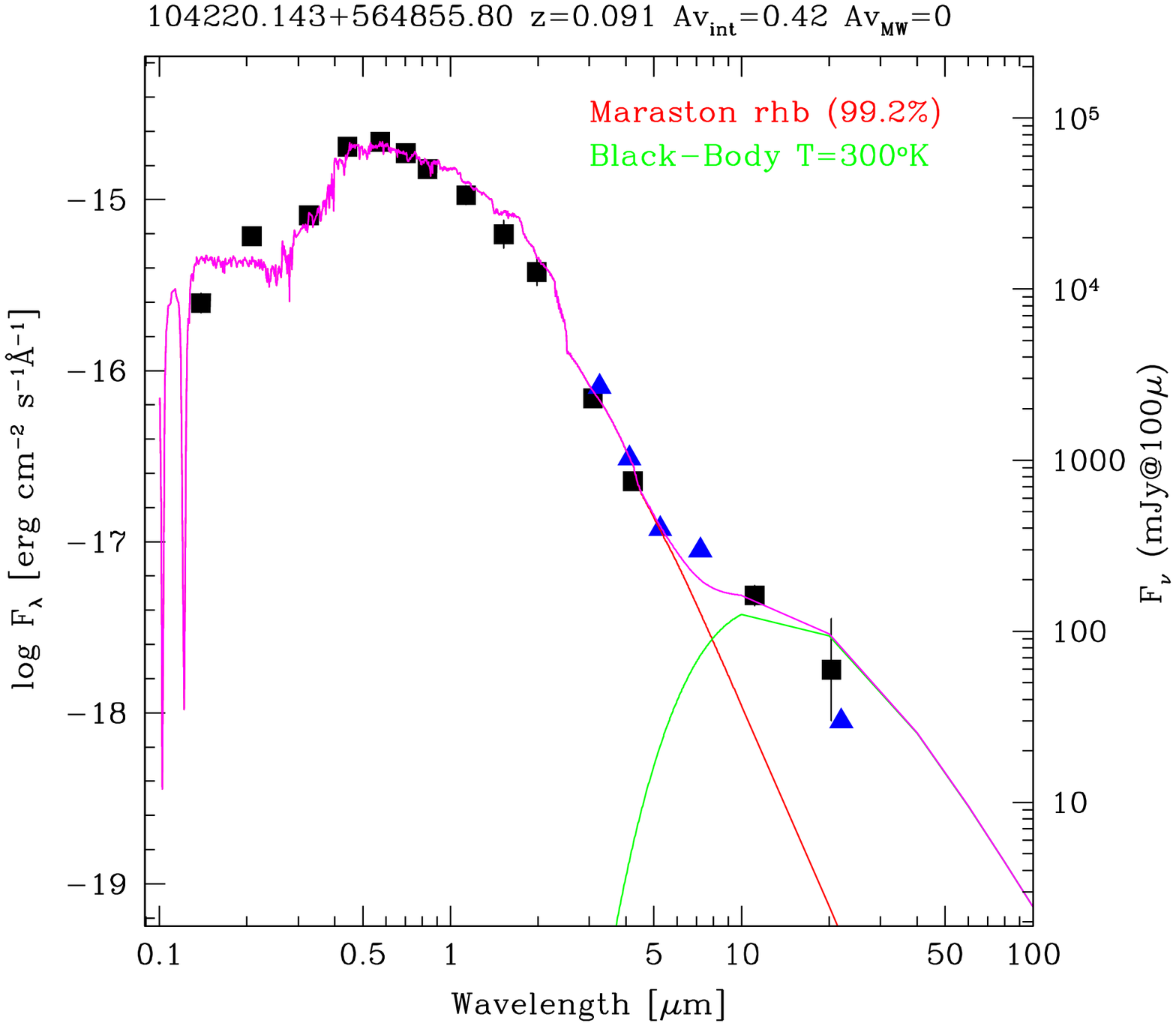}\includegraphics[height=9.0cm, width=9.0cm]{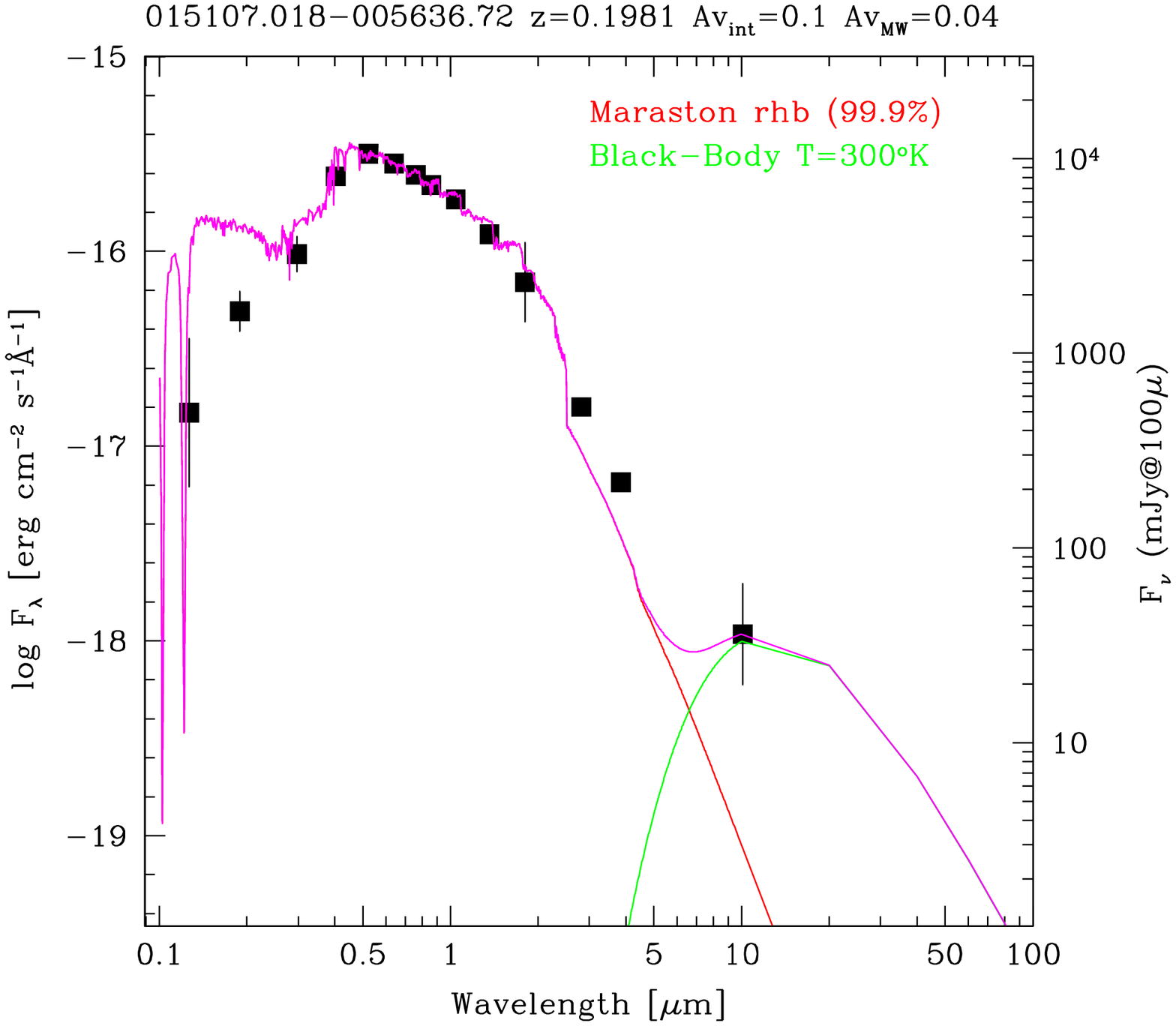}
\caption{{\bf Left.}  SED typical of the K+A galaxies for which we have the complete FUV
to MIR fluxes. The data are seen to be well approximated the {\tt MARASTON-mix} models 
plus a Black-Body of $T_{dust}=300^o$K.  The blue triangles show the SPITZER $3.6\mu$, $4.5\mu$, $5.8\mu$, $8.0\mu$, and $24\mu$ fluxes
for this galaxy, which are seen to match within the errors the corresponding WISE fluxes.
{\bf Right. } Example of an SED that 
is not well fitted by the model at UV wavelengths. Both examples, however, show features 
that are typical of the whole sample: the {\tt Maraston} models overestimate the near-IR 
fluxes; a single 300K blackbody is imposed on the Rayleigh-Jeans tail of the M05 models 
to account for the rising mid-IR flux .}
\label{sed1}
\end{figure*}

We can now analyze the full UV-optical-infrared spectral energy distributions of K+A
galaxies in the light of the information on their stellar populations provided by the
analysis of their spectra and the WHAN diagram. Fluxes retrieved from the
archives were corrected for foreground Galactic extinction using the A$_V$ values of
\cite{Schlafly2011} (downloaded for each source using NED) and for internal extinction 
using the A$_V$ given by the {\tt STARLIGHT} models. In both cases we used the CCM
\citep{Cardelli1989} extinction law for the Milky Way in view of the high metallicity of our
galaxies.  Figure~\ref{sed1} presents sample SEDs for two of the K+A galaxies with full 
FUV to MIR data.

The {\tt STARLIGHT} models use BC03 but one of our aims in this study is to verify whether
TP-AGB stars are needed to reproduce the spectral energy distributions of `young' galaxies
(whether K+As or objects in the distant universe) and we therefore test our observations
against the SSP models of Maraston (2005, hereafter M05), which explicitly treat this stage
of stellar evolution (see Introduction). We used the M05 SSP models for the Kroupa IMF, 
which provide spectral energy distributions for 6 different metallicities ranging from
$\sim$1/180th to 3.5 times solar (-2.25 to +0.67 dex)\footnote{We did not use the 
[M/H]=$- 2.25$ models, however, in order to maximize overlap with the {\tt STARLIGHT}
models that stop at [M/H]=-1.35.} and two different horizontal-branch (HB) morphologies,
chosen heuristically to account for the second parameter problem (bhb for blue and rhb for red horizontal
branches).  

To simplify the analysis, and because at this stage we are mostly
concerned with the ensemble properties of K+A galaxies rather than the detailed modelling
of individual objects, we averaged together the models into three age bins and three
metallicity bins matching as close as possible the ranges spanned by {\tt STARLIGHT} 
as shown in Table~\ref{mix}.

We then used the stellar mass fractions for each {\tt STARLIGHT} bin to generate our model
SEDs as:

\begin{equation}
 \mathfrak{M}_{\lambda} =\frac{ {\rm A}\sum_{i=1}^{6}S_iM_i(\lambda) }{\sum_{i=1}^{6}S_i}
\end{equation}
where for each galaxy $S_i$ are the {\tt STARLIGHT} mass fractions for the six age and
metallicity bins described in Table~\ref{mix} (including only the intermediate-age and old
populations since the fraction of young stars is negligible);  $M_i(\lambda)$ are the
corresponding {\tt MARASTON} bins defined in Table~\ref{mix}.

Notice that since we are not including the young populations, $\sum{S_i}\leq100\%$ 
although for 95\% of the galaxies the sum is $>96\%$ and only for 4 galaxies is the sum
$<80\%$.  Throughout the paper we will refer to the resulting $\mathfrak{M}_{\lambda}$ as
the {\em\tt MARASTON-mix}  for each galaxy, for which the constant A is fixed matching 
the model to the observed fluxes in the SDSS $r'$-band (0.626$\mu$) at the corresponding redshifts.

Figure~\ref{sed1} shows examples of a very good fit (left) and a rather poor fit (right).  Since
the differences in HB morphology are only relevant for old stars, we chose to use the {\tt rhb}
models that cover the full range of ages.\footnote{some of the {\tt bhb} models are 
tabulated for only two ages 10Gyr and 15Gyr, making them much less convenient to use.}
The legends also indicate the fraction of the total stellar mass that was included in the
composite  model spectra for each galaxy. 

Figure~\ref{sed1} (see also Figs.~\ref{colores} and \ref{ccdcomp}) clearly shows that our {\tt Maraston-mix}
SEDs reproduce quite well the observed SEDs from the FUV to the optical (including 
$z'$),  but deviate systematically towards the red; there is a clear excess in the IR, which we
ascribe (see later) to the treatment of TP-AGB stars, and a significant excess over the expected
blackbody behavior at longer wavelengths, where a hot dust component appears to be needed
to account for the observations. 

Determining the temperature of this hot dust component on the
basis of only the WISE data, however, is rather uncertain so we searched the IRAS and SPITZER data bases to 
extend the wavelengths coverage in the mid- to far-IR, and found SPITZER data for 12 sources in the
$3.6\mu - 24\mu$ range, although only a few of these galaxies have complete wavelength coverage in this range.

The data for one such galaxy are plotted in the left panel of Figure~\ref{sed1} as blue triangles. Using the SPITZER data we found that
$T_{dust}\sim300^o$K provides a good fit to the observations of these 12 galaxies, and we therefore adopted the same temperature
for the hot dust component of all the galaxies in our sample, although for individual cases, such as the one shown in the figure, a different
temperature (in this case $\sim$400K) would have provided a better fit.  The uncertainties in the Maraston models discussed below, and the fact that only slightly more that 25\% of our galaxies are detected at 12 and 22 microns, however, 
preclude leaving $T_{dust}$ as a free parameter.

\section{Discussion}

Analysis of the spectra of K+A galaxies shows that these objects appear to consist of  originally
late-type galaxies that have experienced a starburst involving a large fraction of their
stellar mass. We have modeled their stellar populations on the basis of these spectral
fits and we can now compare with the observed SEDs to test models for the origin of
the UV flux, the presence of TP-AGB stars, and the contribution from hot dust.

We assume that the stellar populations deduced from the SDSS spectral fibers can be
extrapolated to the galaxies as a whole. Only integral field spectroscopy can resolve this
difficulty, but it is not generally available for our objects. It seems reasonable to assume 
that the SDSS spectral fibers are well centered on the brightest parts of our relatively 
compact galaxies, but we cannot exclude the possibility of radial population gradients
\citep{Pracy2012}. However, a significant increase in the young stellar populations in the
outskirts of the galaxies, for example, would radically change the FUV to optical SEDs, which
are otherwise very well reproduced by our {\tt Maraston-mix} (which does not include young
stars).  We are confident, therefore, that the SDSS spectra sample reasonably well the overall
stellar populations in K+A galaxies. Barring significant stellar population gradients, the
only free parameter in our simple SED fitting approach is the zero point offset between the
photometric and spectroscopic fluxes, which we calibrated out by matching the models to the 
SDSS $r'$-band fluxes.

\subsection{Emission-line Properties of K+A Galaxies}

We have seen that the overwhelming majority of K+A galaxies are LINERs (i.e. ``low 
ionization nuclear emission-line regions") originally proposed by \cite{Heckman1980} as 
a distinct class of AGN.  However, LINERs are actually a mixed bag of objects with very
different ionization mechanisms (e.g \citealt{TerMel85}). Thus, some LINERs are genuine 
AGN whereas most are actually powered by stars: they are `fake' AGN.

\begin{figure*}
\hspace*{-0.2cm} \includegraphics[height=9cm,width=9cm]{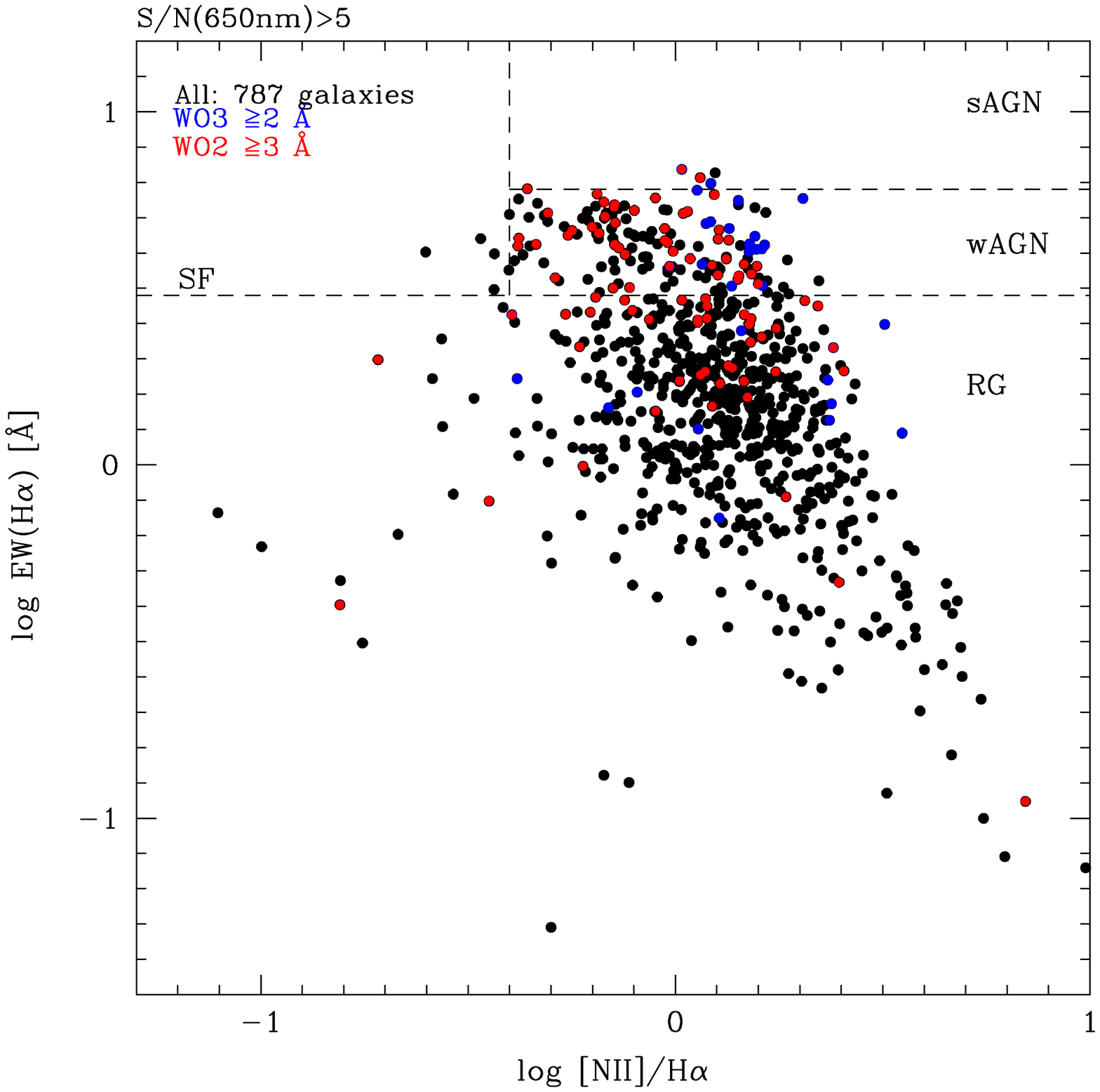}\hspace*{-0.3cm}\includegraphics[height=9cm, width=9cm]{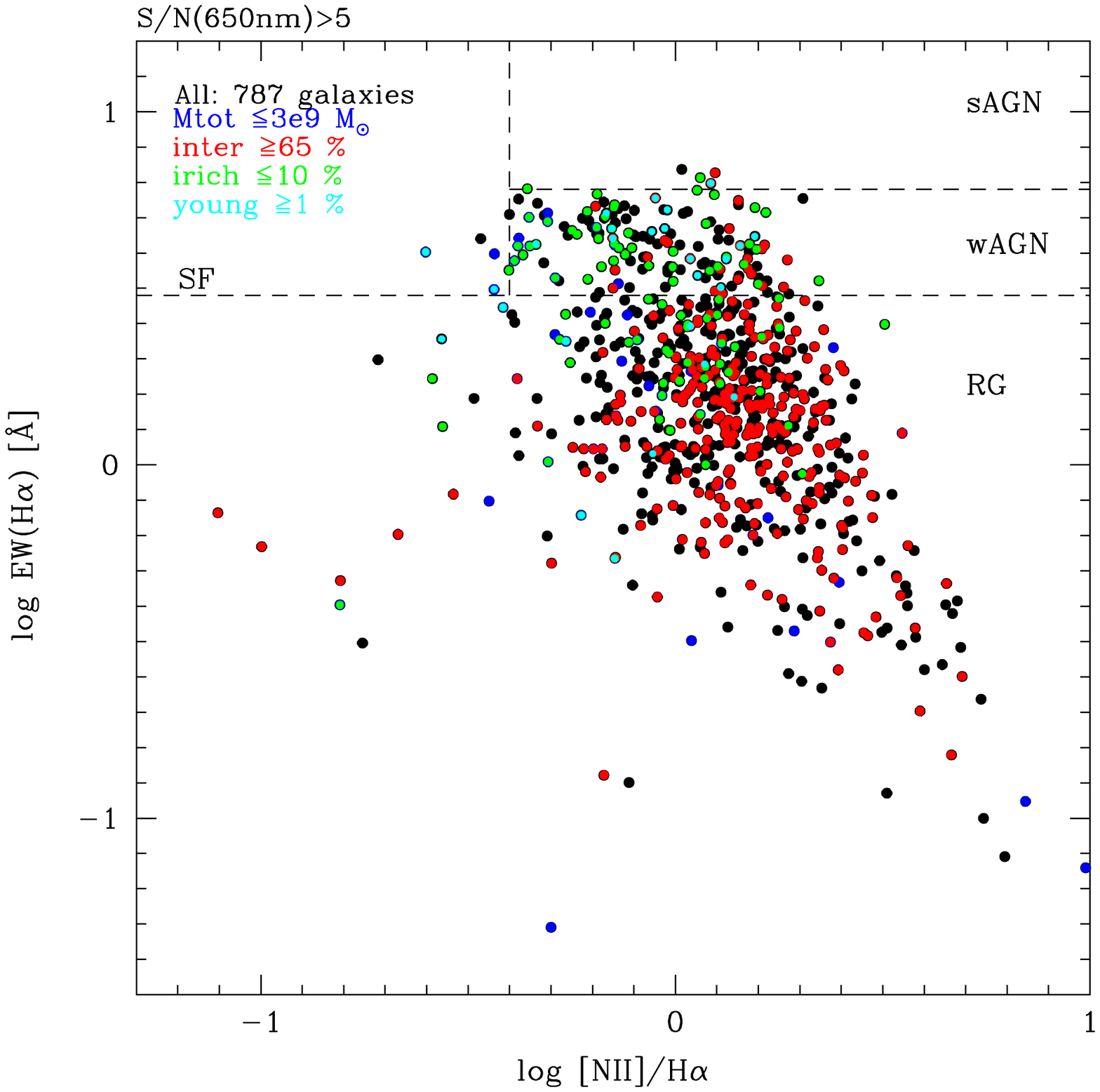}
 \caption{ Same WHAN diagram of Figure~\ref{EWHa} but now using the new classification
scheme proposed by CFS11. Thus, LINERs have been split into `retired galaxies' (RG) and `weak 
AGN' (wAGN), whereas star-forming galaxies (SF) now reside in the upper-left corner 
of the diagram. Strong AGN (sAGN) occupy the region previously ascribed to Seyferts. We 
have used different colors to code the strengths of [OII]3727 and [OIII]5007 in the left panel,
and some relevant parameters from the {\tt STARLIGHT} modes as indicated in the figure captions.  wAGN are
seen to have predominantly metal poor intermediate-age stars; lower masses; higher extinctions; and discernible young populations.  }
 \label{holmes}
\end{figure*}

CFS11 analyzed the spectra of a sample of several hundred thousand galaxies from the 
SDSS, more than half of which turned out to be LINERs. This is by far the most comprehensive
study of the properties of LINERs we are aware of and contains an excellent up to date 
review of the `LINER problem'. On the basis of extensive modelling, CFS11 showed that 
`true' and `fake' LINERs can be nicely separated using their WHAN diagram, and therefore
proposed a new classification scheme for emission-line galaxies. This new scheme is 
presented in Figure~\ref{holmes} that plots the WHAN diagram of K+A galaxies as in
 Figure~\ref{EWHa}.

The new `physically inspired' classification scheme of CFS11 divides emission-line
galaxies into 4 groups as follows: they separate
LINERS into two groups, `weak AGN' (wAGN), and `retired galaxies' (RG). Thus, wAGN are
objects with $\rm 0.6>EW(H\alpha)>0.3$\AA\  and $\rm log([NII]/H\alpha)>-0.4$, which
CFS11 claim to be powered by {\it bona-fide} nuclear black-holes. Conversely, RGs are 
objects with EW(H$\alpha)$$<$0.3\AA\  that are not powered by a nuclear engine, but by 
`hot old low-mass evolved stars' (HOLMES) and therefore were actually `fake' AGNs in the 
old classification scheme. As expected, the galaxies with strongest [OII] \& [OIII] lines lie 
closest to the AGN boundary as also do galaxies with `significant' (i.e. $>$1.5\%) young 
stellar populations. In K+A galaxies the strong UV emission from the intermediate age
populations may supply the UV flux which (in the HOLMES case) is actually provided 
by old Helium-burning stars, since the older stellar populations are not very massive 
in our galaxies.

Although some AGNs are known to exist in K+A galaxies \citep{Liu2007}, Shin, Strauss
\& Tojeiro (2011) find that there are only a few {\it bona fide} AGNs associated with these
galaxies and that the AGN is not related to the previous star formation episode. In our
previous work, we showed that most K+A galaxies do not show evidence of AGN activity
by stacking their 20cm signal \citep{Nielsen2012}. \cite{Brown2009} argue that AGNs
are only responsible for quenching the more massive K+A galaxies, although
\cite{Snyder2011} presents a model where K+A galaxies are the product of gas-rich
mergers, where the K+A phase follows quenching by central AGNs (see also the discussion
in Mendel et al. 2012).

It is not obvious to us that most, or even some, of the objects in the wAGN zone are actually
not `fake' AGN. In order to probe further into this issue in the left panel of 
Figure~\ref{holmes} we  show with different colors galaxies with particularly `strong' 
(in this context!) [OII] and [OIII] lines. These `strong' lined galaxies tend to concentrate 
in the wAGN zone, but many are also RGs, so we can't use these lines as criteria to 
distinguish true from fake AGNs.

In the right-hand panel of Figure~\ref{holmes} we have color-coded the data according to 
some relevant parameters from the {\tt STARLIGHT} models.  The figure clearly shows 
that galaxies where the intermediate-age stars are predominantly metal-poor concentrate 
in the wAGN zone, while galaxies where the intermediate-age populations are metal rich 
tend to avoid that zone and concentrate in the region of RGs.  Galaxies with lower stellar
masses ($\rm M_{tot}$, higher internal extinction, and the few galaxies with significant 
(i.e. $>$1\%!) young stellar populations also tend to concentrate in the SF or wAGN 
domains. The above argues that the engine of the weak emission lines comes of the 
UV emission from intermediate age populations, rather than HOLMES as in other RGs 
studied by CFS11

{\it A priori}, if wAGN are ionized by a central engine, we would not expect to find their 
emission-line properties to be correlated in any way with the stellar populations of their 
parent galaxies, in particular in the case of K+A galaxies where the populations of young 
stars are negligible. But we do see a correlation, which leads us to posit that the majority 
of the K+A galaxies in the wAGN zone may actually be fake AGN. 

The strong excess at $\lambda > 5\mu$m observed in all K+A galaxies (e.g.
Figures~\ref{sed1} and \ref{ccdcomp}) imply that they have substantial amounts of hot 
dust, although we do not know where this dust component is physically located within 
the galaxies (see below for further discussion). The {\tt STARLIGHT} extinctions tell us 
that the dust appears to be well mixed with the gas, and that the intermediate-age 
population appears to be located in regions of lower dust content.  This is consistent
with the model of \cite{Poggianti2000} where A stars migrate out of natal dust clouds,
and the simulations of \cite{Snyder2011} where the intermediate age population is
revealed as dust is cleared by the AGN episode. We will return to this issue in the 
following sections when we analyze the ensemble properties of the dust as told by 
the SEDs.
 
\subsection{Gas-phase metallicities}

Before we leave the emission-line spectra of K+A galaxies, it seems relevant to examine 
whether the metallicities of the nebular component are similar or different to those of
the dominant stellar populations.

\begin{figure}
 \hspace*{-0.0cm}\includegraphics[width=0.5\textwidth]{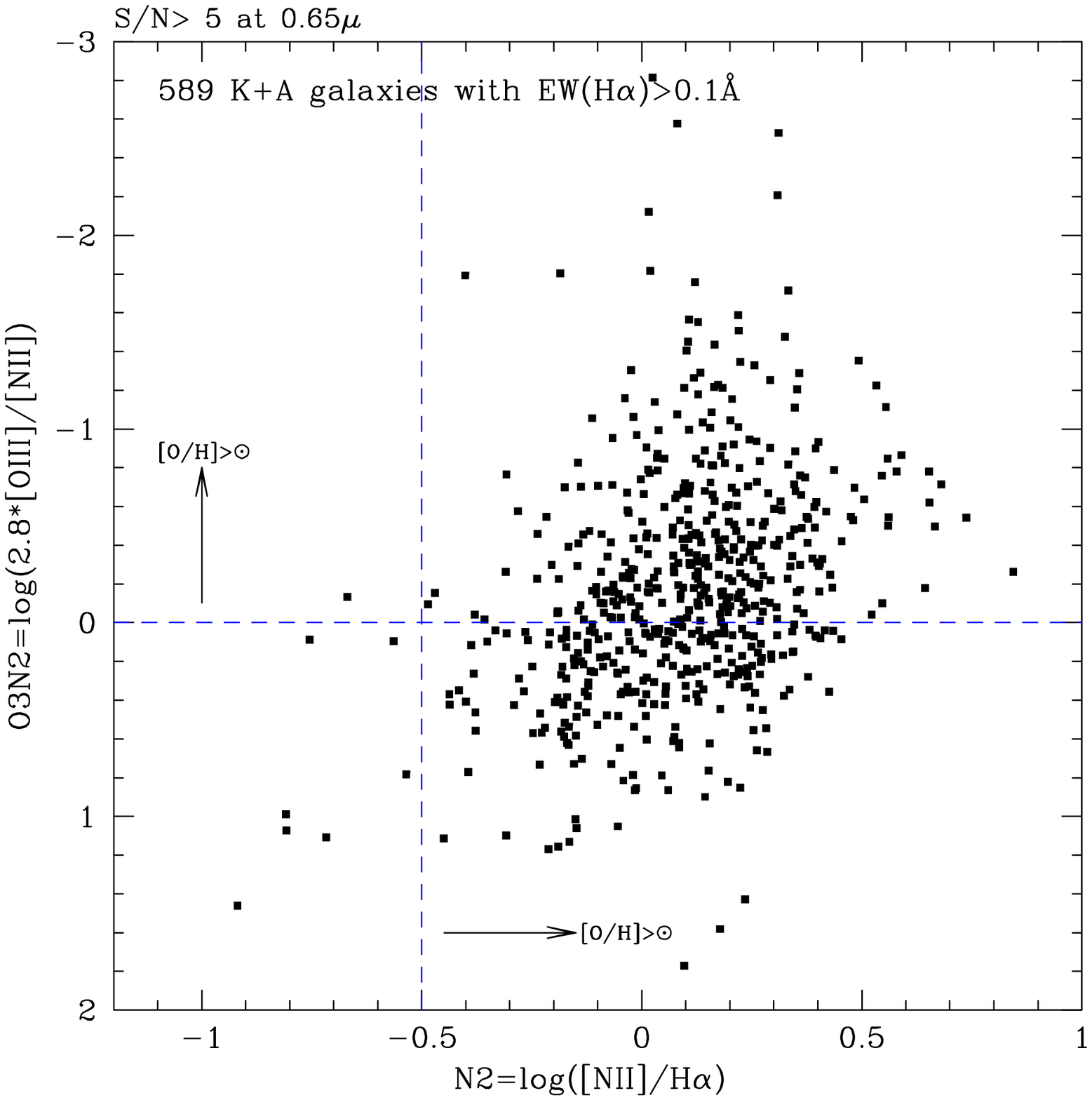}\\
\includegraphics[width=0.5\textwidth]{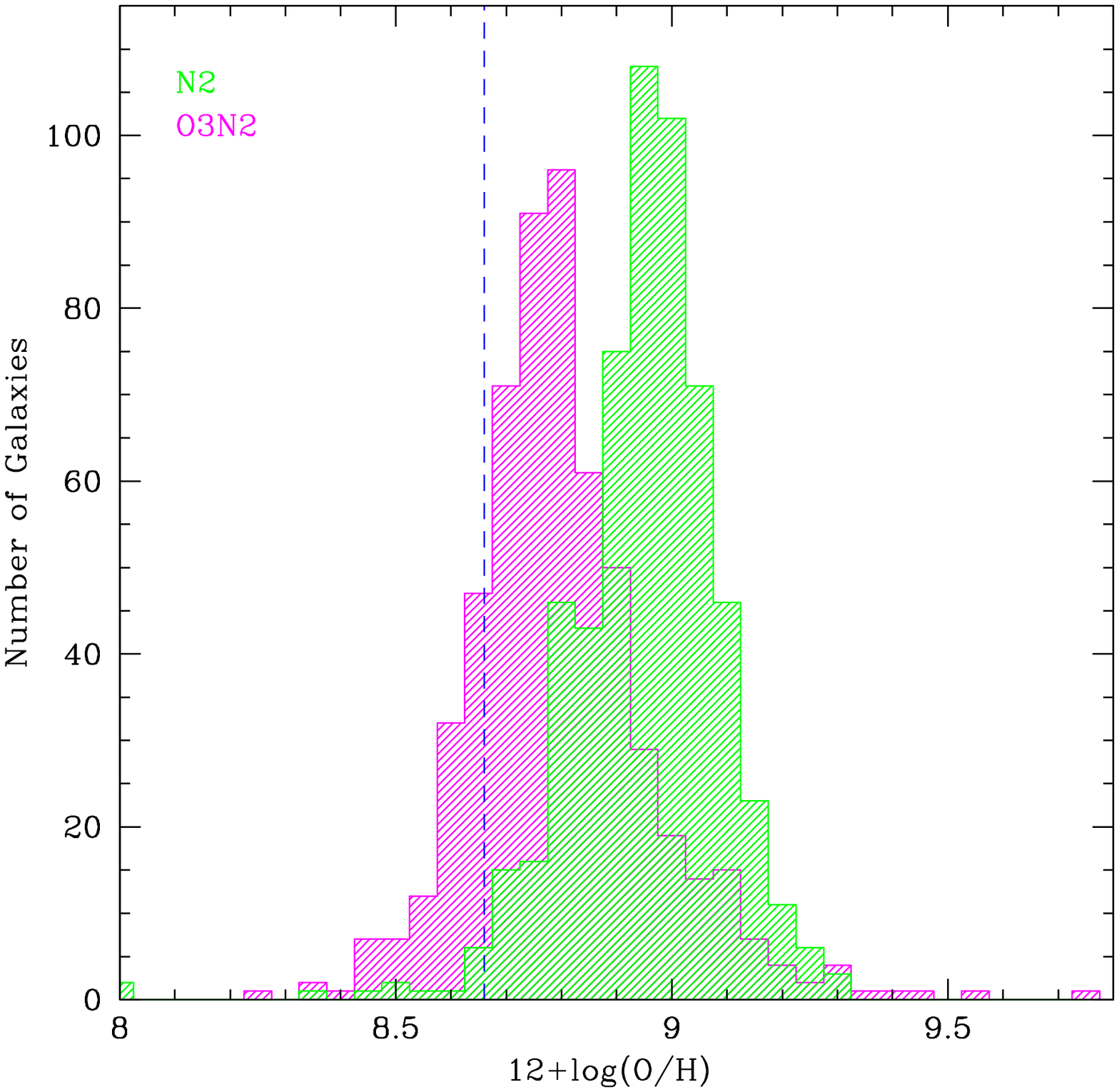} 
 \caption{Gas-phase abundances of K+A galaxies. The top panel shows a scatter plot of the two indices used as proxies for nebular metallicitiy: N2=log([NII/H$\alpha$) and O3N2=log([OIII]/[NII]). The lower panel shows the resulting abundance distributions using the best empirical calibrations from the literature. The dashed blue line shows the solar abundance of oxygen 12+log(O/H)=8.66.}
  \label{r23}
\end{figure}

There are several semi-empirical methods to estimate the metallicities of HII regions
for cases (such as ours) where the physical conditions of the ionized gas, and in particular 
the electron temperatures ($T_e$), are not well known. The most popular of these methods 
is the  so-called  $R_{23}$ originally introduced by \cite{Pagel1979} and later developed 
by  numerous studies (see e.g. \citealt{Peimbert2012} for a recent review).  This method, 
however, is not well suited for our objects because it is ill-determined for metal rich 
nebulae, and because it requires the  [OII]$\lambda$3727,29 fluxes
($R_{23}$=([OII]+[OIII])/H$\beta$), which lie in the noisiest parts of our K+A spectra.

Fortunately, other empirical methods using the more benign [OIII]$\lambda$5007 
and [NII]$\lambda$6584 lines have been extensively studied in the literature (e.g.
\citealt{PP04} henceforth PP04  and references therein), which are much better suited for 
our purposes, in particular because they are calibrated on a `larger' (but still $n<$10) 
sample  of metal-rich nebulae.  Their calibration of these two proxies for metallicity,
$$\rm N2 = [NII]\lambda6584/H\alpha $$
$$\rm O3N2 = \frac{\rm [OIII]\lambda5007/H\beta}{\rm [NII]\lambda6584/H\alpha}$$\\
together with an extensive review of the literature are presented by PP04 and we refer to
that paper for further details. A distinct advantage of both methods is that they are
independent of extinction, although in K+A galaxies, the H$\beta$ nebular line is heavily
contaminated by the stellar absorption from the A-stars, making it difficult to detect, and 
rather uncertain when actually detected. We are therefore forced to use H$\alpha$ instead, 
and hence to correct the observed fluxes for extinction.

This issue has been extensively discussed by CFS11, who derived an empirical relation 
between the nebular extinction, A$\rm^{neb}_V$, estimated from the Balmer decrements 
in  their sample of hundreds of thousands of objects, and the extinction of 
the stellar spectra A$\rm^{*}_V$ resulting from their {\tt STARLIGHT} models. CFS11 remark, 
as we do here, that their empirical relation between A$\rm^{neb}_V$  and A$\rm^{*}_V$
(probably) overestimates the nebular extinction, which provided them with a suitable upper
limit to the parameter space they were interested in exploring. 

The opposite is true in our case where we are interested in a conservative {\em lower limit} for
O3N2, which in any case does  not depend very strongly on extinction.   We therefore used 
the A$\rm ^{*}_V$ values from {\tt STARLIGHT} (without attempting to estimate the 
nebular component) to correct the observed fluxes for extinction.  Following CFS11, we
assumed Case~B recombination for the Balmer decrement (H$\alpha$/H$\beta$=2.8), 
thus, $$\rm O3N2=log(2.8\times[OIII]/[NII]) + 0.12\times A^{*}_V$$ for the CCM 
extinction law.  Figure~\ref{r23} plots  in the top panel O3N2 vs. N2 for those galaxies in 
our sample with S/N$>$5 in the continuum around H$\alpha$, and EW(H$\alpha)>0.1$
(chosen to minimize contamination by spurious features). The dashed lines shows the limits
of $\rm N2> -0.5$  and $\rm O3N2< 0$ corresponding to nebulae with oxygen 
abundances 12+log(O/H) larger than solar  (from Figs. 1\&2 of PP04). PP04 also provide
analytical fits to estimate abundances as,

\smallskip
$$ \rm 12+log(O/H) = 8.90+0.57\times N2 $$  and,
$$ \rm 12+log(O/H) = 8.73-0.32\times O3N2$$
shown in the bottom panel of Figure~\ref{r23}. Within the uncertainties discussed above, the overwhelming majority of our K+A galaxies are
seen to have nebular oxygen abundances well above the solar value (12+log(O/H)=8.66; P04).

Notwithstanding the known discrepancy between the gas and stellar abundances in the
Orion region (which is now narrowed to less than 0.1 dex -- \citealt{Simon2011}), we are
confident that  the ionized gas and the intermediate-age stars in K+A galaxies have 
similar abundance distributions. 


\subsection{UV fluxes}

\begin{figure}
\hspace*{-0.0cm}\includegraphics[width=0.5\textwidth]
{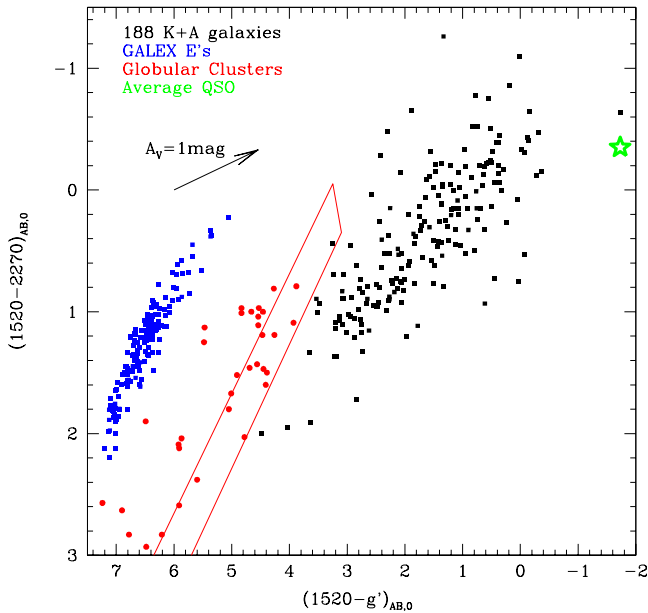}
 \caption{UV color-color diagram for 188 K+A galaxies with both FUV and NUV data from
GALEX, corrected for foreground extinction, plotted in black.  The blue points illustrate the locus of elliptical galaxies in the Coma
cluster, and the red lines delineate the area occupied by horizontal-branch (HB) stars in 
Galactic globular clusters. The red points show integrated GALEX photometry for Galactic 
GCs. The big green star shows the average colors of low-redshift QSOs and the arrow shows the effect of one magnitude 
of extinction for the CCM reddening law. }
 \label{ccda}
\end{figure}

Rising far-UV flux in early-type galaxies has been known to exist since the 1970s with the 
first space UV satellites, and it is now commonly attributed to the presence of hot horizontal
branch stars (see O'Connell 1999 for a review).  Figure~\ref{ccda} presents the  colour-colour
diagram commonly used in the literature to capture the broad statistical properties of the
ultraviolet SEDs of stars and galaxies.  The blue points illustrate the locus of  elliptical galaxies
in Coma \citep{Smith2012}.  These data have not been corrected for foreground extinction, 
which for A$\rm_V=1$mag and the CCM extinction law is represented by the arrow (the
foreground extinction to Coma is of course much less than one magnitude).  

The red lines delimit the area occupied by HB stars in Galactic globular clusters (GC), adapted
from the IUE-based diagram of \cite{Oconnell1999} using the GALEX observations of individual HB 
stars by \cite{Schiavon2012} to transform from IUE to GALEX, and the standard 
(g'-V) color of blue stars. The red points show the integrated colours of a sample of
Galactic GCs from \cite{Daless2012}.  Finally, the green star shows the average colours from 
a sample of 1600 SDSS QSOs at $z\leq0.4$ for which we downloaded the relevant data from the GALEX and SDSS archives.

\begin{figure}
\includegraphics[width=0.5\textwidth]{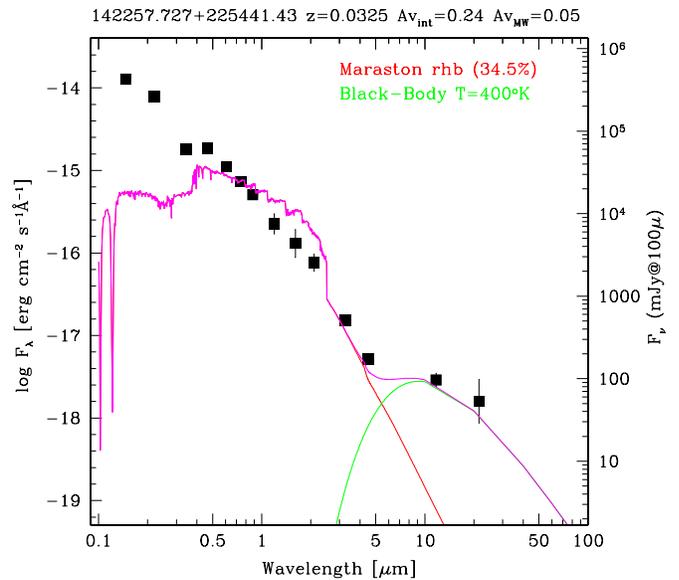} 
\caption{SED of one of the few K+A galaxy in our sample that has bluer FUV colors than typical 
QSOs. This also the only one in our sample with a significant young stellar population 
(35\% of the mass) and lies well in the SF region in the WHAN diagram of 
Figure~\ref{EWHa}. However, it's SED is remarkably similar to a typical QSO.}
 \label{ccdb}
\end{figure}

K+A galaxies are not only substantially bluer than passive ellipticals, as expected, but 
they are also significantly bluer than HB stars in GCs, consistent with the vacuum UV 
fluxes emerging from $\sim 1$ Gyr old stars \citep{Kaviraj2008}. K+A's are however
not as blue as QSO's  (green star), which are more than one magnitude bluer than even the 
bluest galaxies in our sample. There is one galaxy, SDSSJ142257.727+225441.43, however,
that has UV colors typical of QSOs. This is actually one of the few
galaxies in our sample that falls in the SF region of the WHAN diagram (right panel of
Figure~\ref{holmes}) and the only one with a significant population of young stars (39.5\%).
This is also one of the few galaxies in the sample with oxygen abundances lower than solar
(Figure~\ref{r23}), fully consistent with its {\tt STARLIGHT} populations that are dominated 
by young and intermediate-age metal poor stars, with no traces of an old stellar component.
\cite{VeronCetty2006}  classify it as a QSO based on the data by \cite{Darling1994}, but 
this object shows no broad emission lines and no X-ray detection or radio signal in the 
FIRST survey. Interestingly, its SED shown in Figure~\ref{ccdb} is surprisingly similar to 
that of average radio-quiet QSOs (e.g. \cite{Andreani2003}), with a prominent mid-IR 
excess at $10-20\mu$ and a power-law continuum in the blue.
  
\subsection{TP-AGB Stars}
  
\cite{Maraston2005} has suggested that the inclusion of TP-AGB stars in modelling 
intermediate age populations may dramatically alter their colours, derived ages, and 
stellar masses. Testing this hypothesis has proved difficult, as these stars contribute 
most of their light in the infrared that is not easily accessible at high redshifts. However,
\cite{Tonini2009} and \cite{Tonini2010} claim that models including TP-AGB stars 
produce a better match to the colours of distant galaxies, while yielding ages and 
stellar masses in better  agreement with the predictions of theoretical expectations. 
Milner, Rose \& Cecil (2011) obtain a good fit to the (CMD-calibrated) ages of M32 
and NGC5102 using indices derived from infrared spectra and models including 
TP-AGB stars. \cite{Eminian2008} suggest that the infrared colours of star-forming 
galaxies in the SDSS are better modeled by the \cite{Maraston2005} models including 
the TP-AGB contribution. \cite{Melbourne2012} claim that TP-AGB stars may 
contribute as much as 17\% of the $K$-band flux in nearby galaxies. However, 
\cite{Kriek2010} find no evidence that TP-AGB stars are important in their sample 
of higher redshift post-starburst galaxies, while \cite{Zibetti2012} searched for the 
spectroscopic indices of TP-AGB stars in 16 z$\sim$0.2 K+A galaxies and find no 
strong evidence for their presence. The models by \cite{Conroy2009} are in better 
agreement with the older models by \cite{BC03} and yield a lower contribution from 
such stars.

\begin{figure}
\hspace*{-0.0cm}\includegraphics[width=0.5\textwidth]{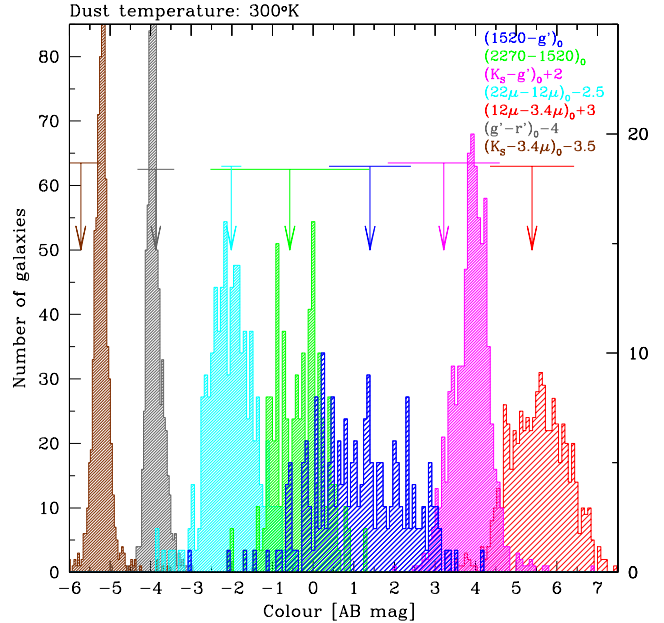}
\caption{ Distribution of observed colors of K+A galaxies using two different scales. The
scale to the left corresponds to the visible to 3.4$\mu$ colors while the scale to the right corresponds
to the colours involving the FUV or 22$\mu$ fluxes. The arrows show the average values for the {\tt MARASTON-mix}+300K BB
fits,  and the dispersion of the sample indicated by the $\pm1\sigma$  bars on top 
of each arrow. A finer binning was used for ($g'-r'$) and ($K_s-3.4\mu$) to avoid compressing the scale of the plot too much.}
\label{colores}
\end{figure}

Figure~\ref{colores} displays some of the broad-band colors commonly used in the 
literature to characterize the SEDs of galaxies. (Notice that several of these colors have 
been shifted as indicated in the figure legend to minimize overlaps.) For each color we 
also plot the average color predicted by our best-fit {\tt  Maraston-mix} plus the 300K
blackbody (which as discussed above we impose {\it ad hoc} to reproduce the observed mid-IR fluxes) shown 
by the arrows, and the dispersion of these predictions plotted as the $\pm1\sigma$
horizontal bars on top of each arrow.

\begin{figure}
\includegraphics[width=0.5\textwidth]{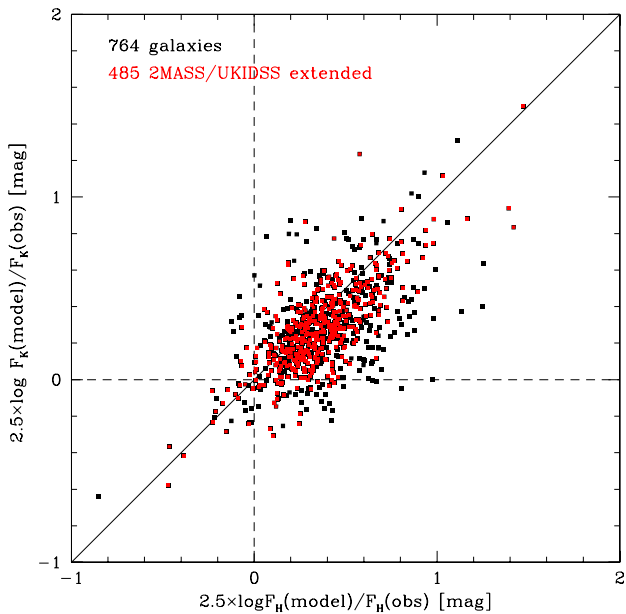} 
\caption{Ratios of predicted to observed near-IR fluxes in the  H and Ks bands of 2MASS and UKIDS. The red symbols represent galaxies for which 2MASS isophotal and/or UKIDSS Petrosian magnitudes are available. The complete overlap of red and black dots shows that the observed ratios are not affected by aperture matching effects. The line for x=y indicates that the models overestimate H more than K.}
\label{irex}
\end{figure}

The M05 models are seen in the figure to reproduce remarkably well
the FUV to optical colors of K+A galaxies. This demonstrates that the FUV colors of 
K+A galaxies are bluer than HB stars because the intermediate-age (A) stars
are actually quite luminous in the UV.   However,  the models systematically predict bluer 
$K_s-g'$ and $K_s-3.4\mu$ colors than observed.  Figure~\ref{irex} shows that these discrepancies actually arise 
because the models overestimate the fluxes in the NIR bands. 

As discussed in Section~\ref{section2}, we are confident that our photometric sample is not affected by systematic aperture matching effects. Nevertheless, we have argued that the influence of TP-AGB stars in the near-IR fluxes of high redshift galaxies is a contentious issue, which needs to be treated with extra care. Therefore, in Figure~\ref{irex} we singled out galaxies classified as extended sources in the 2MASS or UKIDSS catalogues, for which we can be sure that the photometric apertures are ideally matched. The figure shows that extended objects are indistinguishable from the rest of the sample proving that the NIR excess of the models is not due to systematics in the photometry.

Surprisingly,  our simplistic {\tt MARASTON-mix}+BB300K models appear to reproduce remarkably well
the mid-iR colors of the sample, indicating that a single-temperature dust component suffices to reproduce the MIR excess in K+A galaxies (but see below).

\begin{figure}
\includegraphics[width=0.5\textwidth]{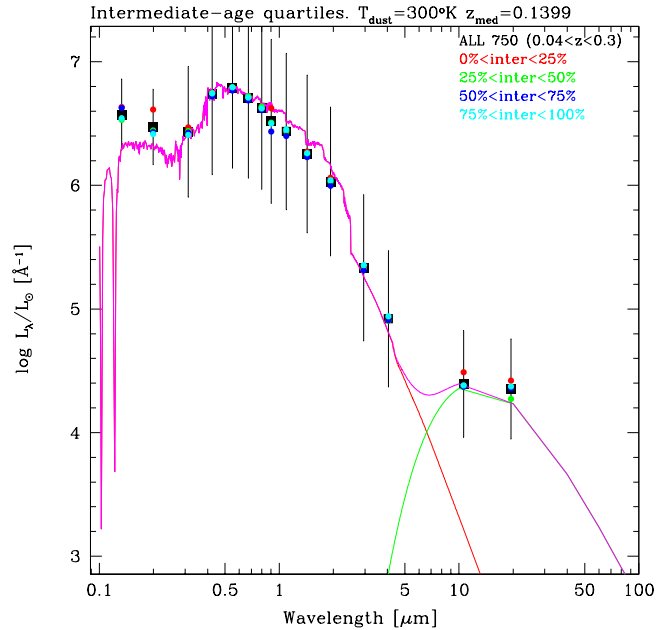}
\caption{Composite SED for K+A galaxies with varying fractions of intermediate age populations (indicated in the figure legend) compared with the normalized  {\it Ensemble \tt MARASTON-mix} obtained by averaging the {\tt STARLIGHT} populations of all the galaxies in the sample fitted to the average fluxes plus a 300K black-body (black squares). In order to optimize the averages we have restricted the redshift range to encompass the bulk of the sample avoiding the extremes.  Notice that the plot is in units of absolute luminosity, so the bars actually indicate the range of luminosities of the galaxies in our sample. Notice that the SEDs do not depend significantly on the intermadiate age fraction.}
\label{ccdcomp}
\end{figure}

Another way of visualizing the systematic properties of the K+A galaxy SEDs is shown in
Figure~\ref{ccdcomp} that presents the averaged SED of K+A galaxies binned in quartiles
of the fraction of intermediate-age stars ({\tt inter}) labelled with different colors. An  {\it Ensemble \tt Maraston-mix}+BB300K,
obtained by averaging the stellar populations of all the galaxies in the sample with redshifts $0.04<z<0.3$, is seen to 
overestimate the observed NIR fluxes but underestimate the 3.4 and 4.6 fluxes. This indicates that  our models may be too simplistic to fully reproduce the SEDs.  

\begin{figure}
  \hspace*{-0.cm}\includegraphics[width=0.5\textwidth]{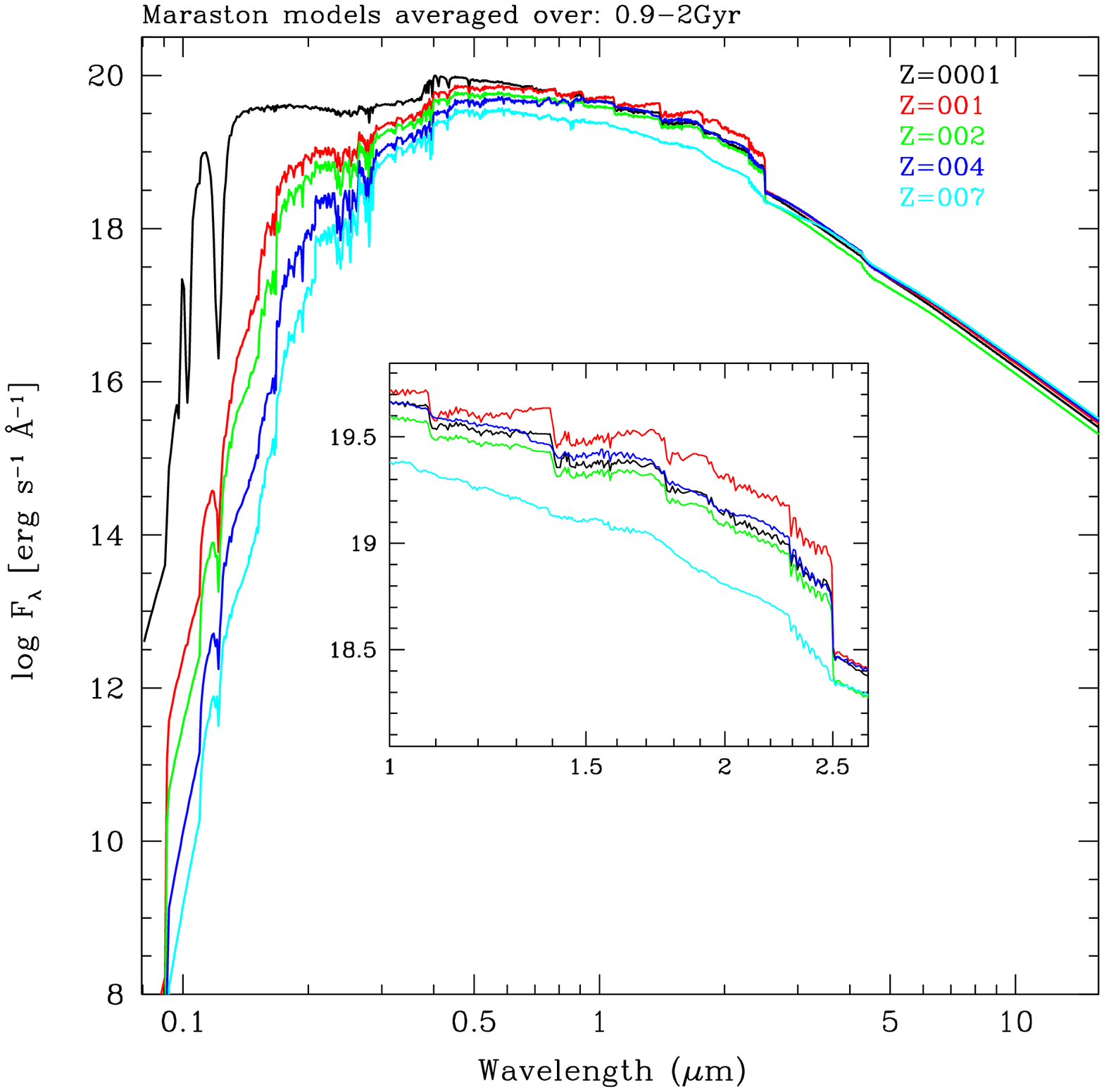}
  \caption{Maraston SSP (Kroupa IMF) models for  intermediate age showing the effect of metallicity, which is dramatic in the UV. The insert magnifies the near IR fluxes between 1 and 2.5$\mu$. }
  \label{models}
\end{figure}

In particular, the discrepancy in the MIR would imply that either the temperature of the hot dust component changes significantly from object-to-object, which would be consistent with the observed dispersion in the WISE colors shown in Fig.\ref{colores}, or that more than one dust component is required to reproduce the observations.  However, some of the trends shown by Figure~\ref{ccdcomp} may actually be due to features of the Maraston models for intermediate age stars. These are illustrated in Figure~\ref{models} that shows the models averaged for stars between $0.9-2$Gyr  used in our {\tt Maraston-mix}, color-coded by metallicity.


The figure shows that the predicted UV fluxes depend dramatically on metallicity such that 
at $\lambda\sim0.15\mu$ metal poor stars are up to six orders of magnitude more
luminous than metal-rich stars.  This would predict that galaxies with the highest fractions 
of (intermediate-age) metal-poor stars should be
systematically over-luminous in the UV.  Since the models also predict  (not shown here) that the UV fluxes of {\em old} metal poor stars
are much larger than those of old metal-rich stars, reaching up to $\sim6$ orders of magnitude at $\lambda\sim0.15\mu$, the UV excess of galaxies with low fractions of intermediate-age stars, represented by the red symbols in Fig.\ref{ccdcomp}, could actually be  the telltale of HOLMES.

In the NIR (magnified in the figure for clarity) the models also show a significant dependence of metallicity. In particular, for $Z$=0.001 and $Z$=0.007, which approximately correspond to our metal-poor and metal-rich bins, the metal-poor model is the most luminous in the NIR and shows the strongest molecular bands, while the metal-rich model is the faintest and shows almost no molecular bands. All the other models (including $Z$=0.0001) lie in between and are roughly similar to one another. 

In the MIR the models  systematically underestimate the observations at 3.4$\mu$ and 4.6$\mu$ (the $12\mu$ point is used to anchor the BB300K component), but the dust temperature cannot be raised too much without compromising the fit at 22$\mu$. However, the models have a discontinuous step at 2.5$\mu$, which is a numerical feature introduced to extrapolate the stellar models into the Rayleigh-Jeans regime at MIR wavelengths. As one would expect, this {\em numerical} step depends on metallicity, although the $Z$=0.001 step is significantly {\em larger} than for $Z$=0.0001. Figure~\ref{models} shows that for stars in our metal-poor bin ($Z\sim0.001$) the 2.5$\mu$ step is about 0.5mag, which corresponds almost exactly to the discrepancy between the {\it Ensemble \tt  Maraston-mix}+BB300K and the observations in the MIR bands.



We must conclude, therefore, that although the models reproduce surprisingly well the
FUV-to-optical colors of K+A galaxies, they provide no evidence that 
TP-AGB stars contribute to the infrared light  in these post-starburst galaxies, in 
agreement with the results of  \cite{Kriek2010} and \cite{Zibetti2012}. The SDSS 
spectroscopy and the broad-band SEDs strongly indicate that the older \cite{BC03} and 
newer \cite{Conroy2009} SSP models are adequate and that an increased contribution 
from TP-AGB stars is not necessary. In turn, this may have important implications for 
the nature of high redshift galaxies, whose colours and luminosities may then truly 
reflect their large ages, stellar masses and metallicities \citep{Yan2004,Cirasuolo2008}.\footnote{The new generation of models by Claudia Maraston (private communication) having a smaller 2.5$\mu$ step will probably provide a better fit in the MIR, but the full models are not yet available for us to check whether they will also provide a better fit in the NIR}

\subsection{FIR excess}

Figure~\ref{ccdcomp} illustrates the fact that {\em all} our galaxies show significant excess 
at MIR wavelengths (in particular $12\mu$ and $22\mu$) compared to the M05 models. 
We have thus added a  Black-Body of $T_{dust}=300^{\circ}$K (BB300K) to describe the 
MIR upturn in the spectra. The temperature of the BB component  
was defined as discussed in Section~\ref{ZEDS} and the normalization fixed by matching the $12\mu$ fluxes as shown in Figure~\ref{sed1}. 
We note here that the cluster objects studied by \cite{Dressler2009} show no significant excess
in the MIPS band at 24$\mu$m, possibly pointing to a real difference between cluster and field
K+A galaxies. On the other hand, 6 out of the 18 galaxies they studied in Abell 851 have
detectable emission at (rest-frame) 16$\mu$m, which is consistent with the fraction
of K+A galaxies we detect in the reddest WISE band.

There are  various possibilities for the source of the mid infrared emission: dusty AGNs, 
dust enshrouded massive stars and clusters, or circumstellar dust from TP-AGB stars
\citep{Kelson2010,Donoso2012,Chisari2012}. The `typical'  $22\mu$-MIR bump 
observed in K+A galaxies is actually quite similar to the average SED of QSOs, as for 
example those by \cite{Andreani2003},  although the FUV up-turn is significantly 
steeper in QSOs. Of course the QSO may be so heavily obscured as to depress the 
UV contribution altogether.

\begin{figure*}
 \includegraphics[width=0.5\textwidth]{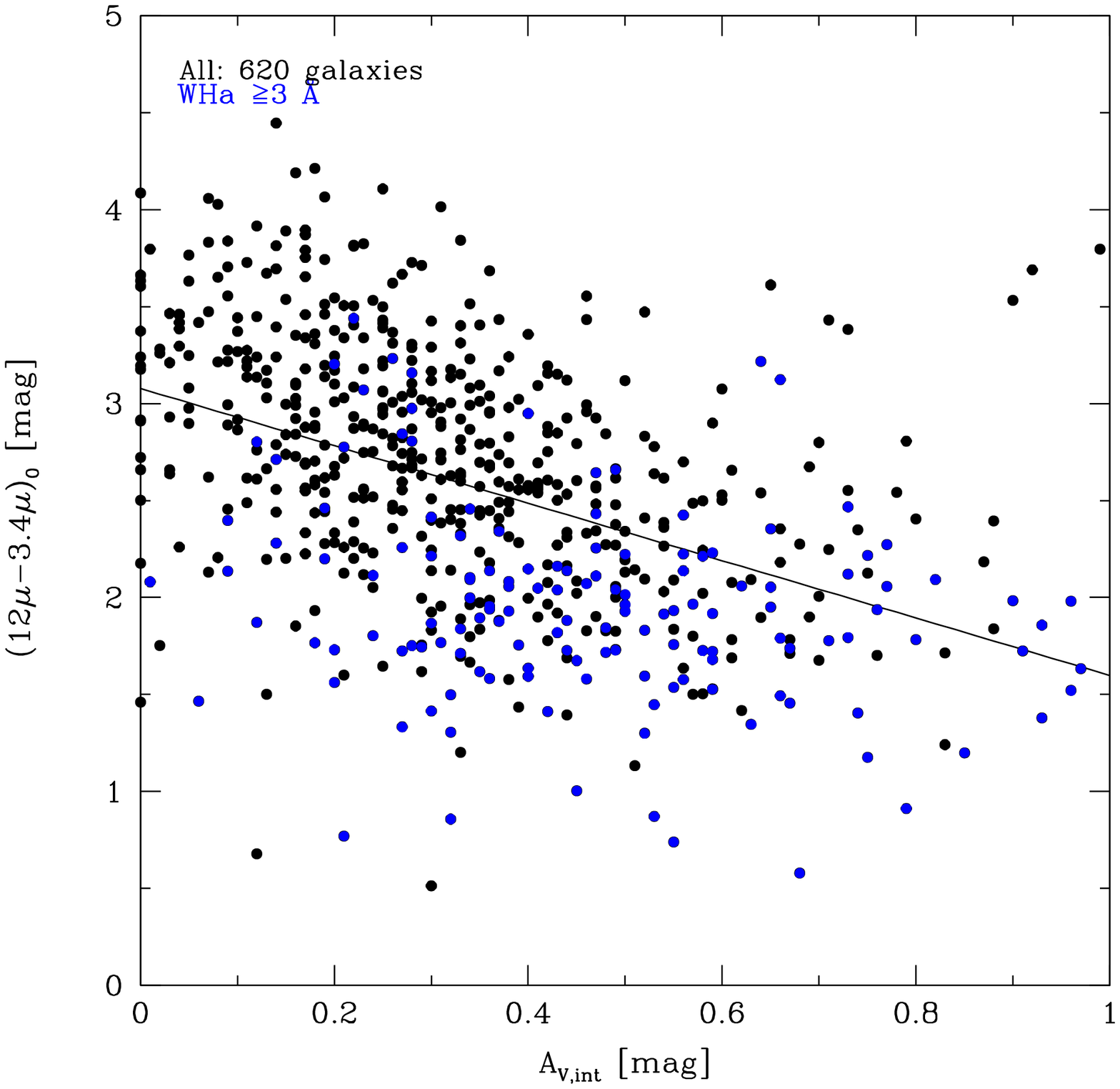}\includegraphics[width=0.5\textwidth]{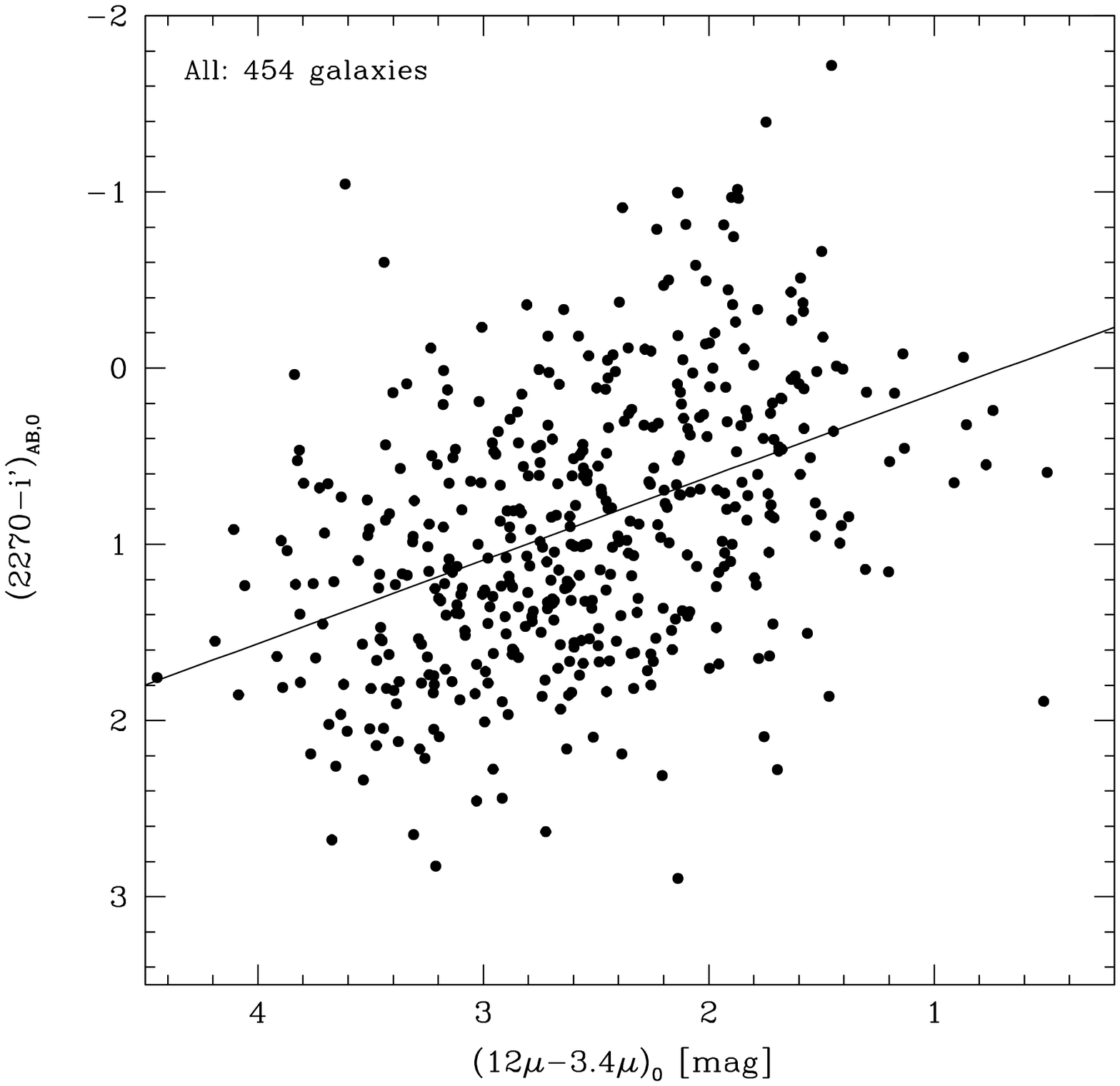} 
 \caption{\small {\bf Left.} Relation between infrared excess and the internal extinction in the galaxies derived from the {\tt STARLIGHT} models showing a clear trend of 12$\mu$ fluxes increasing with extinction.  This indicates that the dust component causing the IR excess is related to the dominant stellar populations. The blue symbols show that for a given amount of extinction, wAGN have systematically larger MIR excesses  {\bf Right.} UV excess plotted as a function of the  $(12\mu-3.4\mu)$ color, which for a single dust component is a proxy of dust temperature. Bluer colors indicate larger UV excess and higher dust temperatures.}
  \label{dust1}
\end{figure*}

To gain further insight into the source of dust heating in K+A galaxies, Figure~\ref{dust1} shows the MIR 
excess index $(12\mu-3.4\mu)$ (which is also a measure of the dust temperature) plotted as a function of internal extinction, $A_{V,int}$
given by {\tt STARLIGHT}. Note that we also used the {\tt STARLIGHT} $A_V$ to de-redden the observed fluxes, although we also verified that our
main conclusions do not depend on extinction.


The figure shows a significant correlation between the MIR excess and the extinction seen 
by the stellar populations. The slope of the correlation is -1.58$\pm$0.006, so the trend 
is highly statistically significant.  The blue points show the galaxies with EW(H$\alpha)>
$3\AA, which corresponds (except for just a few SF objects) to the locus of wAGN galaxies.
Thus, for a given A$_V$  wAGN have more MIR excess than RG.  This may be
 interpreted as evidence that wAGN are enshrouded in dust that is {\bf not seen by the
 dominant stellar populations}, and which must be rather hot in order to radiate at 
$12\mu$.

An important feature of Figure~\ref{ccdcomp} is that on average (black squares) K+A galaxies show a sizable FUV excess relative
to the models; The right panel of Figure~\ref{dust1} shows that the UV excess, measured by ($2270-r'$),  is actually correlated with
the temperature of the hot dust component, measured by 
$(12\mu-3.4\mu)$.\footnote{The correlation is actually stronger for ($1520-g'$) albeit for a much lower number of galaxies}

\cite{Donoso2012} argue that excess flux at 12 $\mu$m may originate from young stars 
and propose the 4.6-12 color in WISE data as a measure of obscured star formation. 
Rising far and mid-IR fluxes are observed in obscured massive stars \citep{Whelan2011} 
so a plausible interpretation of our results is that K+A galaxies are not truly `dead'  but 
that star formation continues within a dusty starburst. Dusty obscured starbursts have 
been proposed as the origin of K+A galaxies by \cite{Shioya2001}, while limits to 
on-going  star formation have been set by \cite{Smail1999, Miller2001,Goto2004} and
\cite{Nielsen2012}.

On the other hand, dust heated by AGB stars in circumstellar envelopes may also become 
quite hot and might explain a mid/far infrared excess, as suggested by \cite{Kelson2010,
Donoso2012,Chisari2012}. Theoretical understanding of this phenomenon does not allow
us to fit models but the predictions suggest that the emission would peak at the intermediate
ages typical of K+A galaxies. 

 
\section{Summary}
CFS11 \citep{Cid2011}  have commented upon the challenges of conveying the information
from large samples of galaxies in an efficient and comprehensible way. Our sample of `only'
808 K+A galaxies pales compared with the mega-studies undertaken by the SEAGal team, 
but we feel we have faced similar challenges. We have attempted the usual martingales of 
binning the data, or using plots in  technicolor, to convey the ensemble properties of 
K+A galaxies, but we still feel that it would be useful for the reader to recapitulate here 
the most important results.  

In summarizing these results, we will make an ontological leap of faith and generalize the
properties of the 808 galaxies in our sample to K+A galaxies as a whole.


\begin{enumerate}

\item K+A galaxies are significantly bluer in the FUV than Horizontal-Branch stars in 
Globular clusters; The FUV to optical colors of K+A galaxies are well reproduced by the
M05 models  as mostly arising from the intermediate-age (A) component;  

\smallskip
\item The overwhelming majority of K+A galaxies have weak but clearly detectable emission
lines. About 75\% of K+A galaxies are `Retired Galaxies' where the nebular component 
is ionized by either intermediate-age stars, or by hot, old, low-mass, evolved stars (HOLMES 
in the nomenclature of CFS11), or both. The remaining $\sim$15\% fall in the category 
of `weak AGN', considered by CFS11 to be genuine AGN;

\smallskip
\item These wAGN have systematically stronger emission lines ([OII], [OIII], H$\alpha$, 
\& [NII]), higher stellar extinction, lower fractions of intermediate-age metal poor stars, 
higher fractions of old metal poor stars, and higher fractions of young stars;

\smallskip
\item Compared to the M05 models, K+A galaxies do not show the predicted near-IR  
excess contribution from TP-AGB stars. It is likely that inclusion of these objects, at least 
at the levels postulated by the models and required to account for the red colors and 
large masses of high redshift galaxies \citep{Tonini2009,Tonini2010}, is not warranted;

\smallskip
\item At least one hot-dust component is required to reproduce the fluxes at
$\lambda > 5$ $\mu$m.  The MIR excess correlates with the stellar extinction indicating 
that at least part of the dust is mixed with the overall stellar populations. At a fixed 
extinction, wAGN and SF have stronger MIR excess than RG's indicating hotter dust temperatures.

\smallskip
\item On average K+A galaxies show a significant FUV excess compared to the models. The temperature of the hot dust component (as given by the MIR excess)
correlates with the FUV excess, reinforcing our deduction that K+A galaxies host massive dust enshrouded starbursts, which could possibly be the genesis of bulges as 
recently suggested by \cite{Mendel2012}.

\end{enumerate}

\begin{figure}
\includegraphics[width=0.5\textwidth]{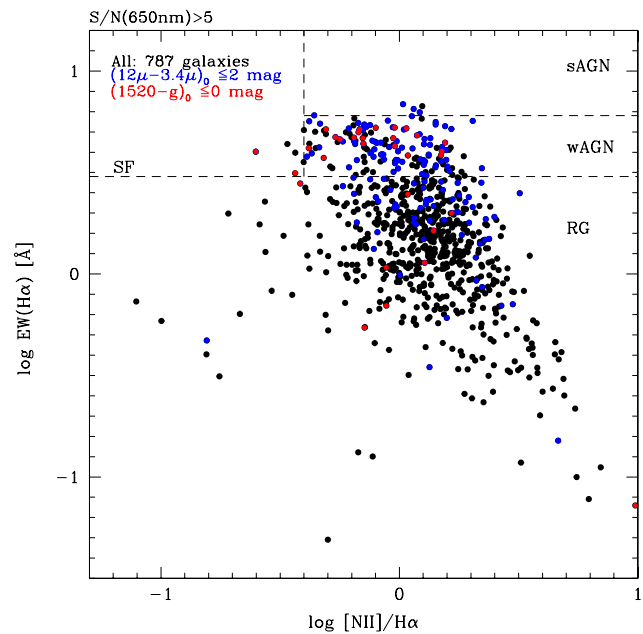}
\caption{WHAN diagram for all 787 galaxies emphasizing objects according to
their $12-3.4$ $\mu$m and FUV-g' colors.}
\label{recap}
\end{figure}

Although in the course of this investigation we have learned a great deal about K+A galaxies,  we have not provided a definitive answer to the fundamental question of whether K+A galaxies are currently forming stars. We found that the signpost of on-going star formation in these objects is not their blue FUV colors, which are mostly due to the intermediate-age (A) stars, but their MIR excess. In particular, the $12\mu$ excess appears to be well correlated with the WHAN type of the galaxies. This is illustrated in Figure~\ref{recap} that shows the WHAN plot color-coded  by the  ($12\mu-3.4\mu$) and ($1520-g'$) colors.  This figure underlines our claim that wAGNs have stronger FUV excesses and hotter dust than RGs, but also, most importantly, that many RGs also share the same properties. 
 
This, and the other `technicolor' figures that we have displayed throughout this paper show that the spectral and photometric properties of K+A galaxies are continuous and not multi-modal, as the WHAN partitioning could lead one to believe. In fact,  a comprehensive review of the literature revealed no significant evidence for non-thermal radio or X-ray emission from K+A galaxies. This would support a model for K+A galaxies as having evolved from 
dusty starbursts in the fashion described by \cite{Shioya2001}, which may be continuing
in highly obscured regions. 

\section*{Acknowledgements}

We are grateful to Roberto Cid-Fernandes, the father of {\tt STARLIGHT}, and to Andr\'e de 
Amorim for proving us with the {\tt STARLIGHT} models for our K+A galaxies as well as 
continuous support in understanding how to use the code. The {\tt STARLIGHT} project is 
supported by the Brazilian agencies CNPq, CAPES and FAPES and by the France-Brazil
CAPES/Cofecub program.
 
This research has made use of the NASA/ IPAC Infrared Science Archive, which is operated by 
the Jet Propulsion Laboratory, California Institute of Technology, under contract with the 
National Aeronautics and Space Administration. 

The initial stages of this publication made use of VOSA \citep{Bayo2008}, developed under the Spanish Virtual
Observatory project supported  from the Spanish MICINN through grant AyA2008-02156. We are indebted to Amelia Bayo for introducing us to VOSA and 
for answering our many questions about VOSA.

The entire GALEX Team gratefully acknowledges NASA's support for construction, operation, and science 
analysis for the GALEX mission, developed in corporation with the Centre National d'Etudes Spatiales of 
France and the Korean Ministry of Science and Technology. We acknowledge the dedicated team of engineers,
technicians, and administrative staff from JPL/Caltech, Orbital Sciences Corporation, University of California, 
Berkeley, Laboratoire d'Astrophysique Marseille, and the other institutions who made this mission possible. 

Funding for the SDSS and SDSS-II has been provided by the Alfred P. Sloan Foundation, the Participating 
Institutions, the National Science Foundation, the U.S. Department of Energy, the National Aeronautics and 
Space Administration, the Japanese Monbukagakusho, the Max Planck Society, and the Higher Education
Funding Council for England. The SDSS Web Site is http://www.sdss.org/.

The SDSS is managed by the Astrophysical Research Consortium for the Participating Institutions. The
Participating Institutions are the American Museum of Natural History, Astrophysical Institute Potsdam, 
University of Basel, University of Cambridge, Case Western Reserve University, University of Chicago, 
Drexel University, Fermilab, the Institute for Advanced Study, the Japan Participation Group, Johns 
Hopkins University, the Joint Institute for Nuclear Astrophysics, the Kavli Institute for Particle Astrophysics 
and Cosmology, the Korean Scientist Group, the Chinese Academy of Sciences (LAMOST), Los Alamos 
National Laboratory, the Max-Planck-Institute for Astronomy (MPIA), the Max-Planck-Institute for Astrophysics 
(MPA), New Mexico State University, Ohio State University, University of Pittsburgh, University of Portsmouth,
Princeton University, the United States Naval Observatory, and the University of Washington.
 
This publication makes use of data products from the Two Micron All Sky Survey, which is a joint project of 
the University of Massachusetts and the Infrared Processing and Analysis Center/California Institute of 
Technology, funded by the National Aeronautics and Space Administration and the National Science 
Foundation

The UKIDSS project is defined in Lawrence et al. (2007). UKIDSS uses the UKIRT Wide 
Field Camera WFCAM (Casali et al. 2007).  The photometric system is described in 
Hewett et al. (2006), and the calibration is described in Hodgkin et al. (2008). The 
pipeline processing and science archive are described in Irwin et al. (2009, in prep) 
and Hambly et al (2008).

This publication makes use of data products from the Wide-field Infrared Survey 
Explorer, which is a joint project of the University of California, Los Angeles, and 
the Jet Propulsion Laboratory/California Institute of Technology, funded by the 
National Aeronautics and Space Administration.

This research has made use of the NASA/IPAC Extragalactic Database (NED) 
which is operated by the Jet Propulsion Laboratory, California Institute of Technology, 
under contract with the National Aeronautics and Space Administration.

\bsp
\label{lastpage}
 

\newpage 
\begin{thebibliography}{99}
\bibitem[\protect\citeauthoryear{Abazajian et al.}{2009}]{Abazajian2009}
Abazajian K. et al. 2009, ApJS, 182, 543
\bibitem[\protect\citeauthoryear{Andreani et al.}{2003}]{Andreani2003} Andreani, P., et al., 2003, AJ, 125, 444. 
\bibitem[\protect\citeauthoryear{Baldry et al.}{2006}]{Baldry2006}
Baldry I. K., Balogh, M. L., Bower, R. G., Glazebrook, K., Nichol R. C., Bamford S. P.,  Budavari T
2006, MNRAS, 373, 469
\bibitem[\protect\citeauthoryear{Bayo et al.}{2008}]{Bayo2008}  Bayo, A., et al.,  2008, A\&A 492, 277
\bibitem[\protect\citeauthoryear{Becker et al.}{1995}]{Becker1995}
Becker R. H., White R. L., Helfand D. J. 1995, ApJ, 450, 559
\bibitem[\protect\citeauthoryear{Bertelli et al.}{1994}]{Bertelli1994} Bertelli G., Bressan A., Chiosi C., Fagotto F., Nasi E., 1994, A\&AS, 106, 275
\bibitem[\protect\citeauthoryear{Brammer et al.}{2009}]{Brammer2009}
Brammer G. et al. 2009, ApJ, 706, L173
\bibitem[\protect\citeauthoryear[{Brown et al.}{2009}]{Brown2009}
Brown M. J. I. et al. 2009, ApJ, 703, 150
\bibitem[\protect\citeauthoryear{Bruzual \& Charlot}{2003}]{BC03} Bruzual, G. \& Charlot S., 2003, MNRAS, 344, 1000
\bibitem[\protect\citeauthoryear{Buyle et al.}{2006}]{Buyle2006}
Buyle P., Michielsen D., De Rijcke S.,  Pisano D. J., Dejonghe H., 
Freeman K. 2006, ApJ, 649, 163
\bibitem[\protect\citeauthoryear{Cardelli, Clayton, \& Mathys}{1989}]{Cardelli1989} 
Cardelli J. A, Clayton G. C., and Mathis J. S., 1989, ApJ,  345, 245 
\bibitem[\protect\citeauthoryear{Casali et al.}{2007}]{Casali2007}
Casali M. et al. 2007, A\&A, 467, 777
\bibitem[\protect\citeauthoryear{ Chabrier}{2003}]{Chabrier2003}
{Chabrier, G, 2003, PASP, 115, 763}
\bibitem[\protect\citeauthoryear{Chang et al.}{2001}]{Chang2001}
Chang T.-C., van Gorkom J., Zabludoff A. I., Zaristsky D., Mihos J. C. 
2001, AJ, 121, 1965
\bibitem[\protect\citeauthoryear{Chisari \& Kelson}{2012}]{Chisari2012}
Chisari N. E., Kelson D. 2012, ApJ, 753, A94
\bibitem[\protect\citeauthoryear{Choi et al.}{2009}]{Choi2009} 
Choi Y., Goto T., Yoon S.-J. 2009, MNRAS, 395, 1776
\bibitem[\protect\citeauthoryear{Cid Fernandes et al.}{2005}]{Cid2005} 
Cid Fernandes R., Mateus A.,  Sodr\'e L., Stasi\'nska G., Gomes J.M.,  2005, MNRAS, 358, 363
\bibitem[\protect\citeauthoryear{Cid Fernandes et al.}{2010}]{Cid2010} 
Cid Fernandes R., Stasi\'nska G., Schlickmann M. S., Mateus A., Vale Asari N., 
Schoenell W., Sodr\'e L. 2010, MNRAS, 403, 1036. 
\bibitem[\protect\citeauthoryear{Cid Fernandes et al.}{2011}]{Cid2011} 
Cid Fernandes R., Stasi\'nska G., Mateus A., Vale Asari N. 2011, MNRAS, 413, 1687. 
\bibitem[\protect\citeauthoryear{Cirasuolo et al.}{2010}]{Cirasuolo2008}
Cirasuolo M., McLure R. J., Dunlop J. S.,  Almaini O., Foucaud S., Simpson C. 2010, MNRAS, 401, 1166
\bibitem[\protect\citeauthoryear{Conroy \& Gunn}{2010}]{Conroy2009}
Conroy C., Gunn J. E. 2010, ApJ, 712, 833
\bibitem[\protect\citeauthoryear{Couch \& Sharples}{1987}]{Couch1987} Couch W. J., Sharples R. M. 
1987, MNRAS, 229, 423
\bibitem[\protect\citeauthoryear{Dalessandro et al.}{2012}]{Daless2012}
Dalessandro, E.,  Schiavon, R.P.,  Rood, R.T.  Ferraro, F.M., Sohn, S.T., Lanzoni, B., 
O'Connell, R.W., 2012, AJ, 144, 126
\bibitem[\protect\citeauthoryear{Darling \& Wegner}{1994}]{Darling1994}
Darling G. W., Wegner G. 1994, AJ, 108, 2025
\bibitem[\protect\citeauthoryear{Denicol{\'o}, Terlevich, \& Terlevich}{2002}]{TerleTerle2002}
Denicol{\'o}, G., Terlevich, R., Terlevich, E.,  2002, MNRAS, 330, 69
\bibitem[\protect\citeauthoryear{Donoso et al.}{2012}]{Donoso2012}
Donoso E. et al. 2012, ApJ, 748, A80
\bibitem[\protect\citeauthoryear{Dressler \& Gunn}{1983}]{Dressler1983} 
Dressler A., Gunn J. E. 1983, ApJ, 270, 7
\bibitem[\protect\citeauthoryear{Dressler et al.}{2009}]{Dressler2009}
Dressler A., Rigby J., Oemler, A., Fritz J., Poggianti B. M.. Rieke G., Bai L. 2009, ApJ, 693, 140
\bibitem[\protect\citeauthoryear{Duc et al.}{2002}]{Duc2002}
Duc P. A. et al. 2002, A\&A, 382, 60
\bibitem[\protect\citeauthoryear{Eminian et al.}{2008}]{Eminian2008}
Eminian C., Kauffmann G., Charlot S., Wild V., Bruzual G., Rettura A., Loveday J. 
2008, MNRAS, 384, 930 
\bibitem[\protect\citeauthoryear{Gonz{\'a}lez Delgado \& Cid Fernandes}{2010}]
 {GonzaCid2010} Gonz{\'a}lez Delgado R. M, Cid Fernandes, R. 2010, MNRAS, 403, 797
\bibitem[\protect\citeauthoryear{Goto}{2004}]{Goto2004}
Goto T. 2004, A\&A, 427, 125
\bibitem[\protect\citeauthoryear{Goto}{2007}]{Goto2007}
Goto T. 2007, MNRAS, 381, 187
\bibitem[\protect\citeauthoryear{Hambly et al.}{2008}]{Hambly2008}
Hambly N. et al. 2008, MNRAS, 384, 637
\bibitem[\protect\citeauthoryear{Heckman}{1980}]{Heckman1980}
Heckman T. M., 1980, A\&A, 87, 142
\bibitem[\protect\citeauthoryear{Hewett et al.}{2006}]{Hewett2006}
Hewett P. C., Warren S. J., Leggett S. K., Hodgkin S. T. 2006, MNRAS, 367, 454
\bibitem[\protect\citeauthoryear{Hodgkin et al.}{2009}]{Hodgkin2009}
Hogdkin S. T., Irwin M. J., Hewett P. C., Warren S. J. 2009, MNRAS, 394, 675
\bibitem[\protect\citeauthoryear{Hopkins et al.}{2008}]{Hopkins2008}
Hopkins P. F., Cox, T. J., Keres D., Hernquist L. 2008, ApJS, 175, 390
\bibitem[\protect\citeauthoryear{Jarosik et al.}{2011}]{Jarosik2011}
Jarosik N. et al. 2011, ApJS, 192, 14
\bibitem[\protect\citeauthoryear{Kaviraj et al.}{2007}]{Kaviraj2007}
Kaviraj S., Kirkby L. A., Silk J., Sarzi M. 2007, MNRAS, 382, 960
\bibitem[\protect\citeauthoryear{Kaviraj et al.}{2008}]{Kaviraj2008}
Kaviraj S. et al. 2008, MNRAS, 388, 67
\bibitem[\protect\citeauthoryear{Kelson \& Holden}{2010}]{Kelson2010}
Kelson D., Holden B. P. 2010, ApJ, 713, L28
\bibitem[\protect\citeauthoryear{Kewley et al.}{2006}]{Kewley2006} 
Kewley, L.J., Groves, B., Kauffnamm, G., Heckman, T., 2006, MNRAS, 372, 961
\bibitem[\protect\citeauthoryear{Kriek et al.}{2010}]{Kriek2010}
Kriek M. et al. 2010, ApJ, 722, L64
\bibitem[\protect\citeauthoryear{Lawrence et al.}{2007}]{Lawrence2007}
Lawrence A. et al. 2007, MNRAS, 379, 1599
\bibitem[\protect\citeauthoryear{Le Borgne et al.}{2003}]{LeBorgne2003}
Le Borgne J.-F. et al. 2003, A\&A, 402, 433
\bibitem[\protect\citeauthoryear[{Liu et al.}{2007}]{Liu2007}
Liu C. T., Hooper E. J., O'Neil K., Thompson D., Wolf M., Lisker T.
2007, ApJ, 658, L249
\bibitem[\protect\citeauthoryear{Maraston}{2005}]{Maraston2005}
Maraston C. 2005, MNRAS, 362, 799
\bibitem[\protect\citeauthoryear{Melbourne et al.}{2012}]{Melbourne2012}
Melbourne J. et al. 2012, ApJ, 748, A47
\bibitem[\protect\citeauthoryear{Melnick et al.}{2012}]{Melnick2012}
Melnick J., Giraud, E., Toledo, I., Selman, Quintana, H., 2012, MNRAS, 427, 47
\bibitem[\protect\citeauthoryear{Mendel et al.}{2012}]{Mendel2012}
Mendel J. T., Simard L., Ellison S. L., Patton D. R. 2012, arXiv, 1211.6115
\bibitem[\protect\citeauthoryear{Miller \& Owen}{2001}]{Miller2001}
Miller N. A., Owen F. N. 2001, ApJ, 554, L25
\bibitem[\protect\citeauthoryear{Milner et al.}{2011}]{Milner2011}
Milner J. , Rose J. A., Cecil G. 2011, ApJ, 727, L15
\bibitem[\protect\citeauthoryear{Morrissey et al.}{2007}]{Morrissey2007}
Morrissey P. et al. 2007, ApJS, 173, 682
\bibitem[\protect\citeauthoryear{Muzzin et al.}{2012}]{Muzzin2012}
Muzzin A. et al. 2012, ApJ, 746, A188
\bibitem[\protect\citeauthoryear{Nielsen et al.}{2012}]{Nielsen2012}
Nielsen D., Ridgway S., De Propris, R., Goto T. 2012, ApJ 761, L16
\bibitem[\protect\citeauthoryear{O'Connell}{1999}]{Oconnell1999} 
O'Connell, R.W., 1999, ARA\&A, 37, 603
\bibitem[\protect\citeauthoryear{Pettini \& Pagel}{2004}]{PP04}
Pettini, M., Pagel, B.E.J., 2004, MNRAS, 348, L59
\bibitem[\protect\citeauthoryear{Pagel et al.}{1979}]{Pagel1979}
Pagel, B. E. J., Edmunds, M. G., Blackwell, D. E., Chun, M. S., Smith, G., 1979, MNRAS, 189, 95
\bibitem[\protect\citeauthoryear{Pe{\~n}a, Peimbert, \& Peimbert}{2012}]{Peimbert2012}
Pen{\~n}a-Guerrero, Peimbert, A., Peimbert, M., 2012, ApJ, 756, L14
\bibitem[\protect\citeauthoryear{Poggianti \& Wu}{2000}]{Poggianti2000}
Poggianti B. M, Wu H. 2000, ApJ, 529, 157
\bibitem[\protect\citeauthoryear{Pracy et al.}{2012}]{Pracy2012}
Pracy M. B., Owers M. S., Couch W. J., Kuntschner H., Bekki K., Briggs F., Lah P.,
Zwaan M. 2012, MNRAS, 420, 2232
\bibitem[\protect\citeauthoryear{Quintero et al.}{2004}]{Quintero2004} 
Quintero A. D. et al. 2004, ApJ, 602, 190
\bibitem[\protect\citeauthoryear{Saintonge et al.}{2008}]{Saintonge2008}
Saintonge A., Tran K.-V. H., Holden B. P. 2008, ApJ, 685, L113
\bibitem[\protect\citeauthoryear{Schiavon et al.}{2012}]{Schiavon2012} 
Schiavon, R.P., et al., 2012,  AJ, 143, 121.
\bibitem[\protect\citeauthoryear{Schlafly \& Finkelbeiner}{2011}]{Schlafly2011}
Schlafly, E., Finkelbeiner, O., 2011, ApJ, 737, 103
\bibitem[\protect\citeauthoryear{Shin et al.}{2011}]{Shin2011}
Shin M.-S., Strauss M. A., Tojeiro R. 2011, MNRAS, 410, 1583
\bibitem[\protect\citeauthoryear{Shioya et al.}{2001}]{Shioya2001}
Shioya Y., Bekki K., Couch W. J. 2001, ApJ, 558, 42
\bibitem[\protect\citeauthoryear{Skrutskie et al.}{2006}]{Skrutskie2006}
Skrutskie M. F. et al. 2006, AJ, 131, 1163
\bibitem[\protect\citeauthoryear{Smail et al.}{1999}]{Smail1999}
Smail I., Morrison G., Gray M. E., Owen F. N.,  Ivison R. J., Kneib J.-P.,  Ellis, R. S. 1999, ApJ, 525, 609
\bibitem[\protect\citeauthoryear{Smith, Lucey, \& Carter}{2012}]{Smith2012} 
Smith R.J., Lucey J.R., Carter D., 2012, MNRAS, 421, 2982
\bibitem[\protect\citeauthoryear{Sim{\'o}n-Diaz \& Stasinska}{2011}]{Simon2011}
Sim{\'o}n-Diaz, S, Stasinska, G.,. 2011, A\&A, 526, A48
\bibitem[\protect\citeauthoryear{Snyder et al.}{2011}]{Snyder2011}
Snyder G. F., Cox T. J., Hayward C. C., Hernquist L., Jonsson P. 2011, ApJ, 741, A77
\bibitem[\protect\citeauthoryear{Stasinska et al.}{2008}]{Stasinska2008} 
Stasinska, G., Vale Asari, N., Cid Fernandes, R., Gomes, J. M., Schlickmann, M., Mateus, A., Schoenell, W., Sodr�, L., Jr., 2008, MNRAS, 391, 29L
\bibitem[\protect\citeauthoryear{Strateva et al.}{2001}]{Strateva2001}
Strateva I. et al. 2001, AJ, 122, 1861
\bibitem[\protect\citeauthoryear{Terlevich \& Melnick}{1985}]{TerMel85}
Terlevich, R.J., Melnick, J., 1985, MNRAS, 213, 841
\bibitem[\protect\citeauthoryear{Tonini et al.}{2009}]{Tonini2009}
Tonini C., Maraston C., Devriendt J., Thomas D., Silk J. 2009, MNRAS, 396, L36
\bibitem[\protect\citeauthoryear{Tonini et al.}{2010}]{Tonini2010}
Tonini C., Maraston C., Thomas D., Devriendt J., Silk J. 2010, MNRAS, 403, 1749
\bibitem[\protect\citeauthoryear{Vergani et al.}{2010}]{Vergani2010}
Vergani S. D. et al. 2010, A\&A, 509, A42
\bibitem[\protect\citeauthoryear{Veron-Cetty \& Veron}{2006}]{VeronCetty2006}
Veron-Cetty M.-P., Veron P. 2006, A\&A, 455, 773
\bibitem[\protect\citeauthoryear{Whelan et al.}{2011}]{Whelan2011}
Whelan D. G., Johnson K. E., Whitney B. A., Indebetouw R., Wood K.
2011, ApJ, 729, A111
\bibitem[\protect\citeauthoryear{Wright et al.}{2010}]{Wright2010}
Wright E. L. et al. 2010, AJ, 140, 1868
\bibitem[\protect\citeauthoryear{Yamauchi et al.}{2008}]{Yamauchi2008}
Yamauchi C., Yagi M., Goto T. 2008, MNRAS, 390, 383
\bibitem[\protect\citeauthoryear{Yan et al.}{2004}]{Yan2004}
Yan H. et al. 2004, ApJ, 616, 63
\bibitem[\protect\citeauthoryear{Yang et al.}{2008}]{Yang2008} 
Yang Y., Zabludoff A.,I., Zaritsky D., Mihos, J.C., ApJ, 688, 945
\bibitem[\protect\citeauthoryear{Zibetti et al.}{2012}]{Zibetti2012}
Zibetti S., Gallazzi A., Charlot S., Pierini D., Pasquali A. 2012, arXiv, 1205.4717
\end{thebibliography}
\end{document}